\documentclass[a4paper,12pt]{article}
\pdfoutput=1
\usepackage[a4paper, bindingoffset=0.2in, left=1in, right=1in, top=1in, bottom=1in, footskip=.25in]{geometry}

\usepackage{times}
\usepackage{amsmath,amssymb}
\usepackage{natbib}
\usepackage{epigraph}
\setcitestyle{citesep={;}}
\usepackage{graphicx,afterpage}
\usepackage[final]{pdfpages}
\usepackage{authblk}
\usepackage[nottoc]{tocbibind}
\usepackage{setspace}
\usepackage{bbm}
\usepackage{hyperref}
\usepackage{xcolor}
\hypersetup{
  colorlinks   = true, %Colours links instead of ugly boxes
  urlcolor     = red, %Colour for external hyperlinks
  linkcolor    = blue, %Colour of internal links
  citecolor   = blue %Colour of citations
}
\usepackage[labelsep=period, labelfont=sc]{caption}
\usepackage{tabularx,ragged2e,booktabs}
\usepackage{sectsty}

\usepackage{caption}
\usepackage{lineno}

\usepackage{enumitem}
\title{Multiscale relevance and informative encoding in neuronal spike trains}

\author[1,2,3,5,*]{\small Ryan John Cubero}
\author[2,4]{\small Matteo Marsili} 
\author[1]{\small Yasser Roudi}

\affil[1]{\footnotesize Kavli Institute for Systems Neuroscience and Centre for Neural Computation, Norwegian University of Science and Technology (NTNU), Olav Kyrres gate 9, 7030 Trondheim, Norway}
\affil[2]{\footnotesize The Abdus Salam International Center for Theoretical Physics, Strada Costiera 11, 34151 Trieste, Italy}
\affil[3]{\footnotesize Scuola Internazionale Superiore di Studi Avanzati, Via Bonomea 265, 34136 Trieste, Italy}
\affil[4]{\footnotesize Istituto Nazionale di Fisica Nucleare (INFN), Sezione di Trieste, Italy}
\affil[5]{\footnotesize Present address: IST Austria, Am Campus 1, 3400 Klosterneuburg, Austria}
\affil[*]{\footnotesize Corresponding author: ryanjohn.cubero@ist.ac.at}

\date{19 December 2019}

\begin{document}
\maketitle

\begin{abstract}
Neuronal responses to complex stimuli and tasks can encompass a wide range of time scales.
Understanding these responses requires measures that characterize how the information on these response patterns are represented across multiple temporal resolutions.
In this paper we propose a metric -- which we call multiscale relevance (MSR) -- to capture the dynamical variability of  the activity of single neurons across different time scales.
The MSR is a non-parametric, fully featureless indicator in that it uses only the time stamps of the firing activity without resorting to any {\em a priori} covariate or invoking any specific structure in the tuning curve for neural activity. 
When applied to neural data from the mEC and from the ADn and PoS regions of freely-behaving rodents, we found that neurons having low MSR tend to have low mutual information and low firing sparsity across the correlates that are believed to be encoded by the region of the brain where the recordings were made.
In addition, neurons with high MSR contain significant information on spatial navigation and allow to decode spatial position or head direction as efficiently as those neurons whose firing activity has high mutual information with the covariate to be decoded and significantly better than the set of neurons with high local variations in their interspike intervals.
Given these results, we propose that the MSR can be used as a measure to rank and select neurons for their information content without the need to appeal to any {\em a priori} covariate.
\end{abstract}

\section{Introduction}

\epigraph{These words I wrote in such a way \\ that a stranger does not know, \\
You too, by way of generosity, \\ read them in way that you know}{The Divan of Hafez}

Much of the progress in understanding how the brain processes information has been made by identifying firing patterns of individual neurons that correlate significantly with the variations in the stimuli and the behaviors.
These approaches have led to e.g. the discovery of V1 cells in the primary visual cortex \citep*{hubel1959receptive}, the A1 cells in the auditory cortex \citep*{merzenich1975representation}, the head direction cells in the anterodorsal nucleus (ADn) of the thalamus \citep*{taube1990head,taube1995head}, the place cells in the hippocampus \citep*{o1971hippocampus} and more recently, the grid cells \citep*{hafting2005microstructure} and speed cells \citep*{kropff2015speed} in the medial entorhinal cortex (mEC).
Such and subsequent studies have selected neurons based on imposed structural assumptions on the tuning profile of neurons with respect to an external correlate.

However, the organization of the brain is hardly this simple and intuitive.
For instance, recent developments in understanding the spatial representation in the mEC have taught us that such approaches has its limits:
First, neurons may break the symmetries of the tuning curves when representing navigational information through shearing \citep*{stensola2015shearing}, field-to-field variability or simply by the constraints of the environment \citep*{krupic2015grid}. Second, the same neuron may respond to a combination of different behavioral covariates, such as position, head direction (HD) and speed in spatial navigation \citep*{sargolini2006conjunctive,hardcastle2017multiplexed}. Finally, and most importantly, neurons may encode a particular behavior in ways that are unknown to the experimenter and that are not related to covariates typically used or to {\em a priori} features.

Under such circumstances, one can still make progress by focusing on the temporal structure of neural firing.
Variations present in the spikes offer neurons with a large capacity for information transmission \citep*{stein1967information,rieke1993coding,stein2005neuronal}.
Recently, it has been shown that the variations found in the relative timing of spikes carry relevant and decodable information about the behavioral task, even when the neuron's firing rate did not increase upon stimulus presentation and decision onset \citep*{insanally2019spike}.
These variations, as captured, for example, by metrics describing the interspike intervals, have been shown to be different for functionally distinct neurons in the cortex \citep*{shinomoto2005measure,shinomoto2003differences,shinomoto2009relating} and have been utilized to classify neurons in the subiculum \citep*{sharp1994spatial} and in the mEC \citep*{latuske2015interspike,ebbesen2016cell}.
However, such measures of variations are either very local or hardly take into account the temporal dependencies and time scales of natural stimuli that lead or contribute to the activity of the neurons.

Here, we propose a novel non-parametric, model-free method for characterizing the dynamical variability of neural spikes across different time scales and consequently, for selecting relevant neurons -- i.e. neurons whose response patterns represent information about the task or stimuli -- that {\em does not require knowledge of external correlates}.
This featureless selection is done by identifying neurons that have broad and non-trivial distribution of spike frequencies across a broad range of time scales.
The proposed measure -- called {\em Multiscale Relevance} (MSR) --  allows an experimenter to rank the neurons according to their information content and relevance to the behavior probed in the experiment. The theoretical arguments that lead to the definition of MSR are laid out in a number of recent publications on efficient representations \citep*{marsili2013sampling,haimovici2015criticality,cubero2019criticality}; see also \citep*{battistin2017learning} for a concise review of this theoretical work. These arguments have been shown to be useful in characterizing the efficiency of representations in deep neuronal networks \citep*{song2017emergence} and in Minimum Description Length codes \citep*{cubero2018entropy}, as well as for identifying relevant sites in proteins \citep*{grigolon2016identifying}. The aim of this paper is to show that these arguments can also be used for studying neural representations by applying it to real and synthetic neural data.

We illustrate the method by applying it to data on spatial navigation of freely roaming rodents in  \citet*{stensola2012entorhinal} and \citet*{peyrache2015internally}, that report the neural activities of 65 neurons simultaneously recorded from the medial Entorhinal Cortex (mEC), and of 746 neurons in the Anterodorsal thalamic nucleus (ADn) and Post-Subiculum (PoS), respectively.
In all cases, we find that neurons with low MSR also coincide with those that contain no information on covariates involved in navigation, but that the opposite is not true.
We find that some neurons with high MSR also contain significant information for spatial navigation, some relative to position, some to HD but often on both space and HD.
These findings corroborate the recent conjecture of multiplexed coding \citep*{panzeri2010sensory} both in the mEC \citep*{hardcastle2017multiplexed}, the thalamus \citep*{mease2017multiplexed} and the subiculum \citep*{lederberger2018}. We observe that MSR correlates to different degrees with different measures that have been introduced to characterize specific neurons.
More specifically, we find strong correlation between MSR and measures of sparse representations of external correlates. Furthermore, we show that the neurons in mEC with highest MSR have spike patterns that allow a downstream decoder ``neuron'' to discern the organism's state in the environment. Indeed, the top most relevant neurons (RNs), according to MSR, decode spatial position (or HD) just as well as the top most spatially (or HD) informative neurons (INs). In addition, we find that this decoding efficiency can not solely be due to local variations in the interspike intervals \citep*{shinomoto2005measure,shinomoto2003differences}. Emphasizing again that the MSR does not rely on any information about space or HD and is calculated only from the timing on spikes, the correlation with spatial or HD information suggests a role for MSR as an unsupervised method for focusing on information-rich neurons without knowing {\em a priori} what covariate(s) those neurons represent. 

\section{Multiscale Relevance}

Consider a neuron whose activity is observed up to a time $t_{obs}$. This can be one of a population of $N$ simultaneously recorded neurons in the same experiment. 
%We consider a population composed of $N$ neurons in an animal whose activities were simultaneously observed up to a time, $t_{obs}$.
%From hereon, we shall focus our attention to a single neuron within this population.
The activity of this neuron is recorded and stamped by the spike times $\lbrace t_1, \ldots, t_{M} \rbrace$ where $t_1 < t_2 < \ldots \leq t_{M} \leq t_{obs}$ and $M$ is the total number of observed spikes.
By discretizing the time into $T$ bins of duration $\Delta t$, a spike count code, $\lbrace k_1, k_2, \ldots, k_T \rbrace$, can be constructed where $k_s$ denotes the number of spikes recorded from the neuron in the $s$\textsuperscript{th} time bin $B_s=[(s-1)\Delta t,s\Delta t)$ ($s=1,2,\ldots,T$).

Fixing $\Delta t$ allows us to probe the neural activity at a fixed time scale.
Yet, rather than using $\Delta t$ to measure time resolution, we adopt an information theoretic measure, given by
\begin{equation}
H[s] = -\sum_{s = 1}^T \frac{k_s}{M} \log_M \frac{k_s}{M},
\label{resolution1}
\end{equation}
where $\log_M(\cdot)=\log(\cdot)/\log M$ indicates logarithm base $M$ (in units of $M$ats).
Considering $k_s/M$ as the probability that the neuron fires in the bin $B_s$, this has the form of a Shannon entropy \citep*{cover2012elements}.
This corresponds to the amount of information that one gains on the timing of a randomly chosen spike by knowing the index $s$ of the bin it belongs to\footnote{With no prior knowledge, a spike can be any of the $M$ possible spikes, so its {\em a priori} uncertainty is of $\log_2 M$ bits. The information on which bin $s$ the spike occurs, reduces the number of choices from $M$ to $k_s$ and the uncertainty to $\log_2 k_s$ bits. Averaging the information gain $\log M-\log k_s$ over the {\em a priori} distribution of spikes and dividing by $\log M$, yields Eq. \eqref{resolution1}. It is also worth to stress that Eq. \eqref{resolution1} does not refer to the estimate of the entropy of a hypothetical underlying distribution $p_s$ from which spikes are drawn. This would not make much sense, because it is well-known that the {\em na\"ive} estimate of the Shannon entropy in Eq. \eqref{resolution1} obtained with the maximum likelihood estimator $\hat p_s=k_s/M$ suffers from strong biases \citep*{treves1995upward,strong1998entropy}.}.

We argue that $H[s]$ provides an intrinsic measure of resolution, contrary to $\Delta t$ which refers to particular time scales that may vary across neurons.
For example, there is a value $\Delta t_-$ such that for all $\Delta t \le \Delta t_-$, all time bins either contain a single spike or none, i.e. $k_s=0,1$ for all $s$.
All these values of $\Delta t$ correspond to the same value of the intrinsic resolution $H[s]=1$. 
Likewise, there may be a value $\Delta t_+$ such that for all $\Delta t\ge \Delta t_+$, all spikes of the neuron fall in the same bin. 
All $\Delta t\ge \Delta t_+$ then correspond to the same value $H[{s}]=0$ of the resolution, as defined here. 
In other words, $H[s]$ captures resolution on a scale that is fixed by the available data.

Given a resolution $H[s]$ (corresponding to a given $\Delta t$), we can now turn to characterize the dynamic response of the neuron.
The only way in which the dynamic state of the neuron in bin $s$ can be distinguished from that in bin $s'$ is by its activity.
If the number of spikes in the two bins is the same ($k_s= k_{s'}$) there is no way to distinguish the dynamic state of the neuron in the two bins, at that resolution\footnote{One may argue that, if the activity in the previous bins $s-1$ and $s'-1$ differs considerably, then the dynamic state in bin $s$ and $s'$ may also be considered different. We take the view that this distinction is automatically taken into account when considering larger bins (i.e. $\Delta t\to 2\Delta t$).}; see \cite{cubero2019criticality} for a general argument underlying this statement. Therefore, one way to quantify the richness of the dynamic response of a neuron is to count the number of different dynamic states it undergoes in the course of the experiment. A proxy of this quantity is given by the variability of the spike frequency $k_s$, that again can be measured in terms of an entropy
\begin{equation}
H[K] =  -\sum_{k=1}^M \frac{k m_k}{M} \log_M\frac{k m_k}{M}.
\label{relevance}
\end{equation}
where $m_k$ indicates the number of time bins containing $k$ spikes\footnote{$m_k$ satisfies the obvious relation $\sum_{k=0}^M k m_k = \sum_{s=1}^T k_s = M$.}, so that $k m_k/M$ is the fraction of spikes that fall in bins with $k_s=k$.
Again, rather than considering $H[K]$ as a Shannon entropy of an underlying distribution $p_k\approx km_k/M$ of spike frequencies, we take $H[K]$ as an information theoretic measure of the information each spike contains on the dynamic state of the neuron at a given resolution\footnote{Again, the knowledge of the associated dynamical state, i.e. the spike frequency $k$ of the bin it belongs to, provides information to identify the timing of a spike by reducing the number of possible choices from $M$ to $km_k$, which is the number of spikes in bins with the same dynamical state $k$. The information gain is given by $H[K]$.}.
\citet*{cubero2019criticality} show that $H[K]$ provides an upper bound to the number of informative bits that the data contains on the generative process. Also $H[K]$ correlates with the number of parameters a model would require to describe properly the dataset, without overfitting \citep*{haimovici2015criticality}.
Hence, following \citet*{cubero2019criticality}, we shall call $H[s]$ as {\em resolution} and $H[K]$ as {\em relevance}. 

In the current context, the reason for this choice can be understood as follows.
In a given task or behavior, different neurons can have activities that are more or less related to the behavioral or neuronal states that are being probed in the experiment.
Neurons that are {\em relevant} for encoding the animal's behavior or task are expected to display rich dynamical responses, i.e. to have a large $H[K]$. 
On the contrary, neurons that are not involved in the animal's behavior are expected to visit relatively fewer dynamic states, i.e. to have a lower $H[K]$. 

Notice that for very small binning times $\Delta t \le \Delta t_-$ (when each time bins contains at most one spike, i.e. $m_{k=1}=M$ and $m_{k'}=0, ~\forall~k' > 1$) we find $H[K]=0$ (and $H[s]=1$).
At the opposite extreme, when $\Delta t \ge \Delta t_+$ and $H[s]=0$, we have all spikes in the same bin, i.e. $m_k=0$ for all $k=1,2,\ldots, M-1$ and $m_M=1$. Therefore again we find $H[K]=0$.
Hence, no information on the relevance of the neuron can be extracted at time scales smaller than $\Delta t_-$ or larger than $\Delta t_+$.
At intermediate scales $\Delta t\in [\Delta t_-,\Delta t_+]$, $H[K]$ takes non-zero values that we take as a measure of the relevance of the neuron for the freely-behaving animals being studied, at time scale $\Delta t$. 
\begin{figure*}[!ht]
\begin{center} 
\includegraphics[width=\textwidth]{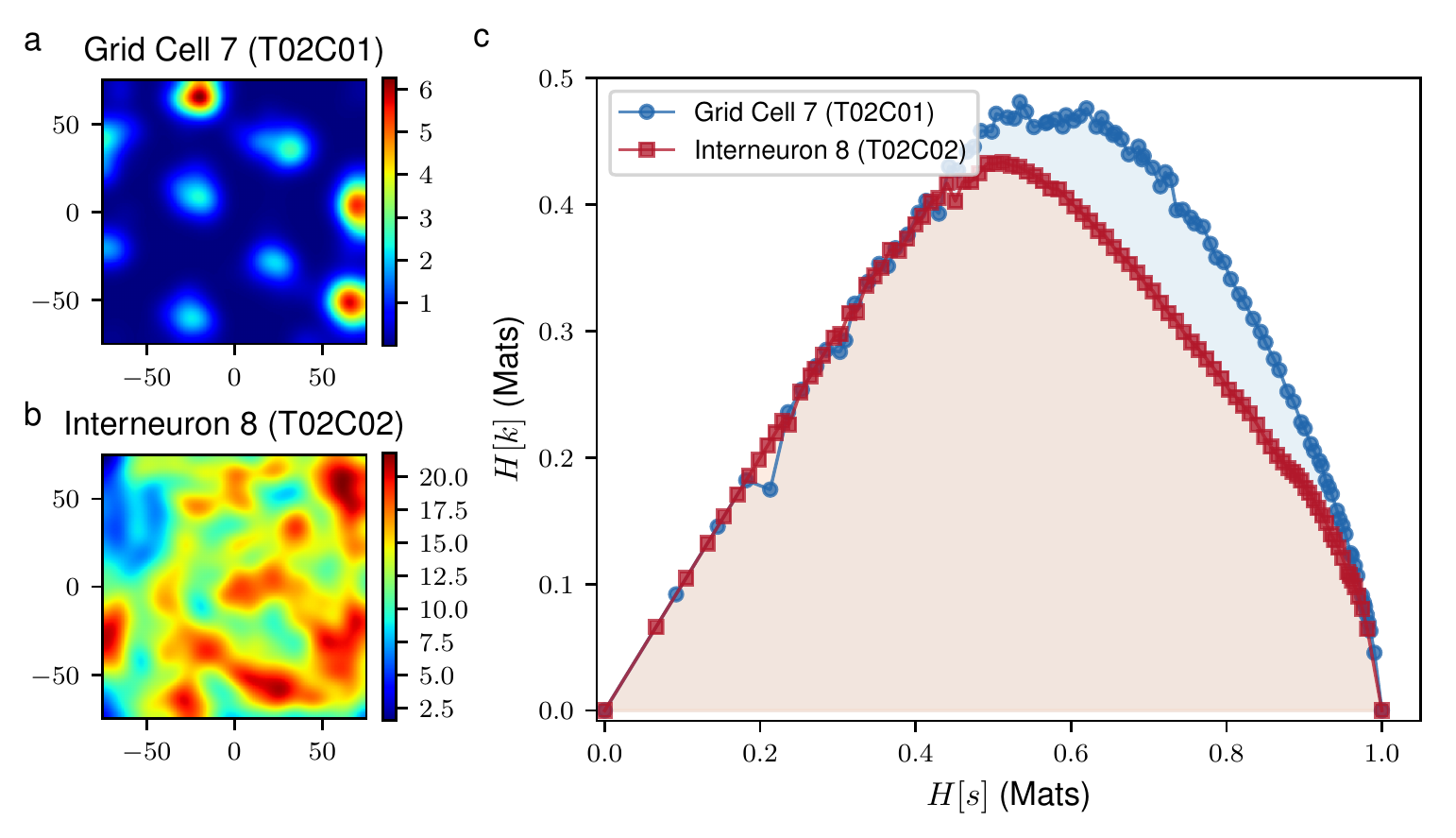}
\end{center} 
\caption{\label{figure Figure1}
{\bf Proof of concept of the MSR as a relative information content measure}.
The smoothed firing rate maps of a grid cell ({\bf a}) and an interneuron ({\bf b}) in the mEC illustrates the spatial modulation of neural activity.
Panel {\bf c} shows the curves traced by the grid cell (blue) and interneuron (red).
Each point, $(H[s],H[K])$, in this curve corresponds to a fixed binning time, $\Delta t$, with which we see the corresponding temporal neural spike codes.
} 
\end{figure*}

Yet, the relevant time scale $\Delta t$ for a neuronal response to a stimulus may not be known {\em a priori} and/or the latter may evoke a dynamic response that spans multiple time scales.
For this reason, we vary the binning time $\Delta t$ thereby inspecting multiple time scales with which we want to see the temporal code.
As we vary $\Delta t$, we can trace a curve in the $H[s]$-$H[K]$ space for every neuron in the sample.
Neurons with broad distributions of spike frequencies across different time scales will trace higher curves in this space and in turn, will cover larger areas under this curve (see Fig. \ref{figure Figure1}c).
Henceforth, we shall call the area under this curve as the \emph{multiscale relevance} (MSR), $\mathcal{R}_t$.
The \emph{relevant neurons} (RNs), those with high values of $\mathcal{R}_t$, are expected to exhibit spiking behaviors that can be well-discriminated by downstream neural information processing units over short and long time scales and thus, are expected to be \emph{relevant} to the encoding of higher representations. On the theoretical note, \citet*{marsili2013sampling} show that, for a given value of $M$ and of the resolution $H[s]$, data that are maximally informative on the generative process are those for which $H[K]$ takes a maximal value. In the high resolution region (small $\Delta t$ or large $H[s]$), the frequency distributions that achieve maximal values of $H[K]$ are broad. More precisely, the frequency distribution behaves as $m_k\sim k^{-\mu-1}$ where $-\mu$ is the slope of the $H[K]-H[s]$ curve. Indeed, $\mu$ quantifies the tradeoff between resolution ($H[s]$) and relevance ($H[K]$) in the sense that a reduction in $H[s]$ of one bit delivers an increase of $\mu$ bits in $H[K]$ \citep*{cubero2019criticality}.

MSR is designed to capture non-trivial structures in the spike train stemming from the variations in spike rates.
As such, it is expected to correlate with other measures characterizing temporal structure, such as bursty-ness and memory \citep*{goh2008burstiness} and the coefficient of local variation in the interspike interval \citep*{shinomoto2005measure,shinomoto2003differences} (see Text S1 for details).
We have observed that, in synthetic data with given characteristics, MSR captures both the bursty-ness and memory of a time series, and local variations in the interspike intervals (see Fig. \ref{figure FigureS1}a,b and Text S1 for definition).
In addition, we find, in both synthetic and real data, a negative relation between MSR and spike frequency (i.e. $M$), which is partly associated with bursty-ness. Finally, we performed extensive tests on synthetically generated time series to show that the MSR captures non-trivial structure induced by the dependence of neural activity on external covariates (see Fig. \ref{figure FigureS1}g and the discussion on Fig. \ref{figure Figure6}). 

As a proof of concept of the MSR for featureless neural selection, we considered two neurons recorded simultaneously from the medial entorhinal cortex (mEC) by \citet*{stensola2012entorhinal} -- a grid cell (T02C01) and an interneuron (T02C02) -- both of which were measured from the same tetrode and thus, are in close proximity in the brain region.
The mEC and its nearby brain regions are notable for neurons that exhibit spatially selective firing (e.g., \emph{grid cells} and \emph{border cells}) which provides the brain with a locational representation of the organism and provides the hippocampus with its main cortical inputs.
Grid cells have spatially selective firing behaviors that form a hexagonal pattern which spans the environment where the rat freely explores as in Fig. \ref{figure Figure1}a.
Apart from spatial information, grid cells can also be attuned to the HD especially in deeper layers of the mEC \citep*{sargolini2006conjunctive}.
These cells altogether provide the organism with an internal map which it then uses for navigation.
On the other hand, interneurons, as in Figure \ref{figure Figure1}b, are inhibitory neurons which are still important towards the formation of grid cell patterns \citep*{couey2013recurrent, pastoll2013feedback, roudi2014grid} but have much less spatially specific firing patterns.
Intuitively, as the mEC functions as a hub for memory and navigation, grid cells, which provide the brain with a representation of space, should be more relevant for a downstream information processing ``neuron'' (possibly the place cells in the hippocampus) in encoding higher representations compared to interneurons.
Indeed, the grid cell traces a higher curve in the $H[s]-H[K]$ diagram of Fig. \ref{figure Figure1}c, thus enclosing a larger area, as compared to the interneuron.

\section{Results}
Following the observations in Fig. \ref{figure Figure1}, we sought to characterize the temporal firing behavior of the 65 neurons which were simultaneously recorded from the mEC and its nearby regions of a freely-behaving rat as it explored a square area of length 150 cm \citep*{stensola2012entorhinal}.
This neural ensemble, as functionally categorized by \citet*{stensola2012entorhinal}, consisted of 23 grid cells, 5 interneurons, 1 putative border cell and 36 unclassified neurons, some of which had highly spatially tuned firing and nearly hexagonal spatial firing patterns \citep*{stensola2012entorhinal,dunn2015correlations,dunn2017grid}.
This dataset was chosen among the multiple recording sessions performed by \citet*{stensola2012entorhinal} as this contained the most grid cells to be simultaneously recorded.

These results were then corroborated by characterizing the temporal firing behaviors of the 746 neurons which were recorded from multiple anterior thalamic nuclei areas, mainly the anterodorsal (AD) nucleus, and subicular areas, mainly the post-subiculum (PoS) of 6 different mice across 31 recording sessions while the mouse explored a rectangular area of dimensions 53 cm $\times$ 46 cm \citep*{peyrache2015internally}.
This data was chosen as these heterogeneous neural ensembles contained a number of \emph{HD cells} which are neurons that are highly attuned to HD.

Before showing the results on these data sets, we note that the the MSR is a robust measure.
To establish this, we compared the MSRs computed using only the first half of the data to that computed from the second half.
We obtained very similar results, confirming that the MSR is a reliable measure that can be used to score neurons (see Fig. \ref{figure FigureS2}a).

\subsection{MSR captures information on functionally relevant external correlates.}
As the mEC is crucial to spatial navigation, we sought to find whether the wide variations of neural firing as captured by the MSR would contribute towards a representation of the animal's spatial organization, in one way or another.
Different measures relating the spatial position, $\mathrm{\mathbf x}$, with neural activity had been employed in the literature to characterize spatially specific neural discharges, like the Skaggs-McNaughton spatial information, $I(s,\mathrm{\mathbf x})$ defined in Eq. \eqref{equation information} and by \citet*{skaggs1993information}, spatial sparsity measure, $sp_{\mathrm{\mathbf x}}$ defined in Eq. \eqref{equation sparsity} and by \citet*{skaggs1996theta} and \citet*{buetfering2014parvalbumin} and grid score, $g$, defined in Eq. \eqref{equation gridscore} and by \citet*{sargolini2006conjunctive}, \citet*{dunn2017grid}, \citet*{solstad2008representation} and \citet*{langston2010development}.

Apart from spatial location, HD also plays a crucial role in spatial navigation. 
The mean vector length, $R$ (Eq. \eqref{equation meanvectorlength} in Section \ref{methods scores}) is commonly used as a measure of HD selectivity of the activity of neurons. 
However, this measure assumes that there is only one preferred HD in which a given neuron is tuned to.
Hence, we calculated two measures -- the HD information, $I(s,\theta)$, and HD sparsity, $sp_\theta$ -- inspired by the spatial information and spatial sparsity to quantify the information and selectivity of neural firing to HD respectively.
These measures ought to detect non-trivial and multimodal HD tuning which may also be important in representing HD in the brain \citep*{hardcastle2017multiplexed}.

\begin{figure*}[!h]
\begin{center}
\includegraphics[width=\textwidth]{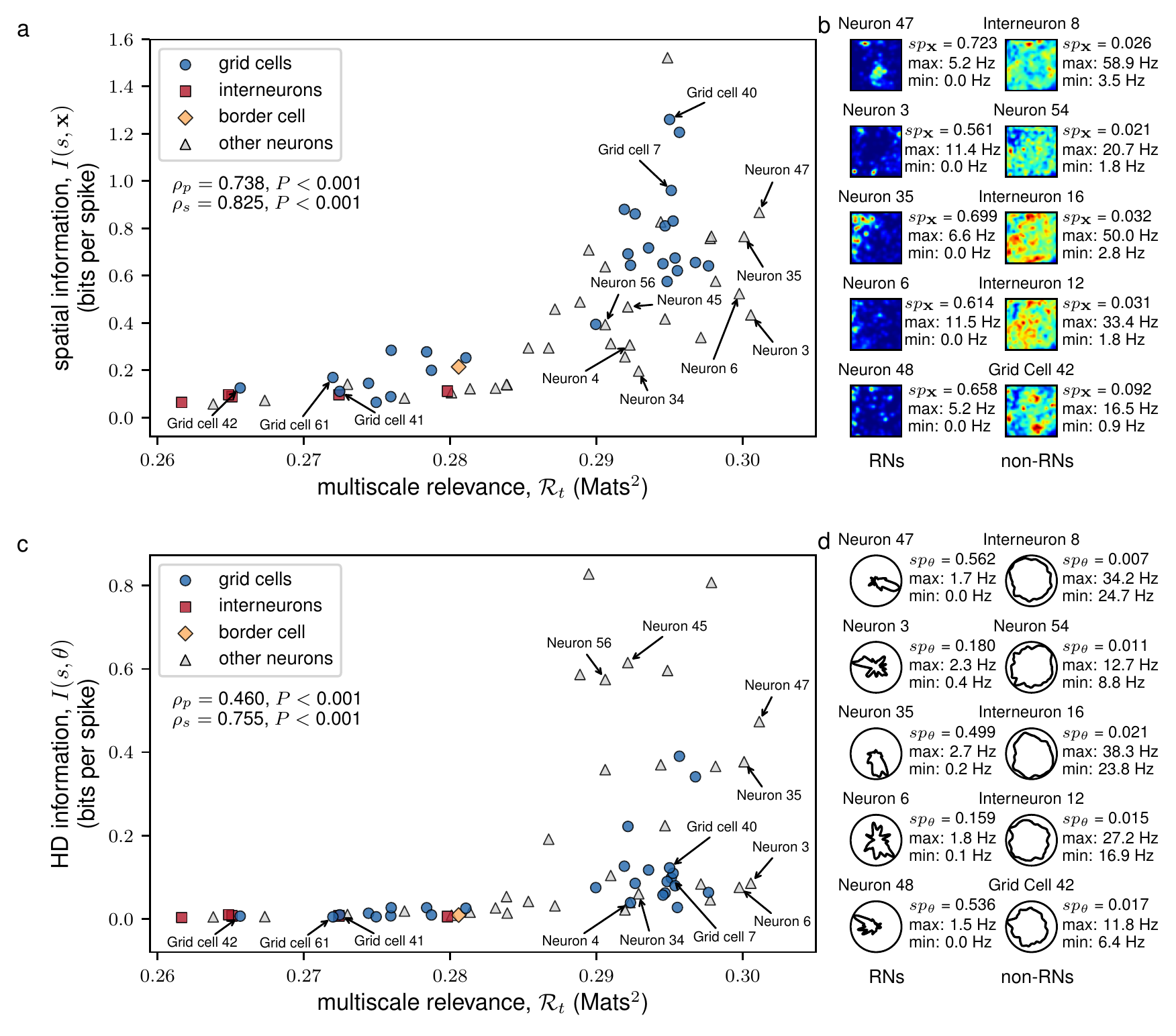}
\end{center} 
\caption{\label{figure Figure2}
{\bf The MSR identified neurons that are spatially and head directionally informative}.
A scatter plot of the MSR vs. the spatial (HD) information is shown in {\bf a} ({\bf c}).
The shapes of the scatter points indicate the identity of the neuron according to \citet*{stensola2012entorhinal}.
The linearity and monotonicity of the multiscale relevance and the information measures were assessed by the Pearson's correlation, $\rho_p$, and the Spearman's correlation, $\rho_s$, respectively.
Information bias was measured by a bootstrapping method, i.e., calculating the average of the spatial or head directional information of 1000 randomized spike trains.
The spatial firing rate maps (HD tuning curves) of the 5 most relevant neurons (RNs) and the 5 most irrelevant neurons (non-RNs) are shown together in panel {\bf b} ({\bf d}) together with the calculated spatial sparsity, $sp_{\mathrm{\mathbf x}}$, (HD sparsity, $sp_\theta$) and maximum and minimum firing.
}
\end{figure*}

Fig. \ref{figure Figure2} reports the spatial information (a) and the HD information (c) as a function of the MSR for each neuron in the mEC data.
Figs. \ref{figure Figure2}b and d report the spatial firing rate maps and HD tuning curves for the top five RNs (left panel) and non-RNs (right panel) by MSR score, respectively (See also Figs. \ref{figure FigureS4} and \ref{figure FigureS5}).
We observed that non-RNs had very non-specific spatial and HD discharges as indicated by their sparsity scores (Figs. \ref{figure Figure2}b and d, Figs. \ref{figure FigureS4} and \ref{figure FigureS5}) whereas RNs had a broader range of spatial and HD sparsity (Figs. \ref{figure Figure2}b and e, Figs. \ref{figure FigureS4} and \ref{figure FigureS5}).

While we have observed that the MSR has a negative relation with the spike frequency (i.e. $M$), an analysis of the residual MSR revealed that the logarithm of the spike frequency (i.e., $\log M$) could not explain all of the variations in the MSR for the neurons in the mEC.
We have seen that the residual MSRs (with respect to $\log M$) appeared to be correlated with spatial and HD information (see Figs. \ref{figure FigureS2}b-d).

Although local variations in the interspike intervals, as measured by $L_V$, could still capture spatial and HD information (see Figs. \ref{figure FigureS3}a and b, respectively), we observed that the strength of correlation was stronger for MSR than for $L_V$.
While there is a positive correlation between $L_V$ and the MSR (see Fig. \ref{figure FigureS2}e), we found that local variations could not explain what is captured by the MSR.
In addition, the residual MSRs (with respect to $L_V$) were observed to still be correlated with spatial or HD information (see Figs. \ref{figure FigureS2}f and g).

We found that {(\em i)} Neurons with high spatial information or high HD information also had high MSR, but the converse was not true.
While there were highly RNs that responded exquisitely to space (grid cells 7 and 40) or HD (neurons 45 and 56) alone, the majority (e.g. neurons 35 and 47) encoded significantly both spatial and HD information. 
Secondly, we found that {\em ii)} Neurons with low MSR had both low spatial and low HD information (Figs. \ref{figure Figure2}b and d, right panel), but again, the converse was not true (e.g. neurons 4 and 34).
Finally {\em iii)} we found that some neurons, for example, neurons 3 and 6, despite having some spatial and HD sparsity as indicated in their rate maps (Figs. \ref{figure Figure2}b and d, left panel), had relatively low spatial and HD information but were both identified to be RNs by MSR.
This high MSR suggests that perhaps these neurons responded to other correlates involved in navigation different from spatial location or HD.

\begin{figure*}[!h]
\begin{center}
\includegraphics[width=\textwidth]{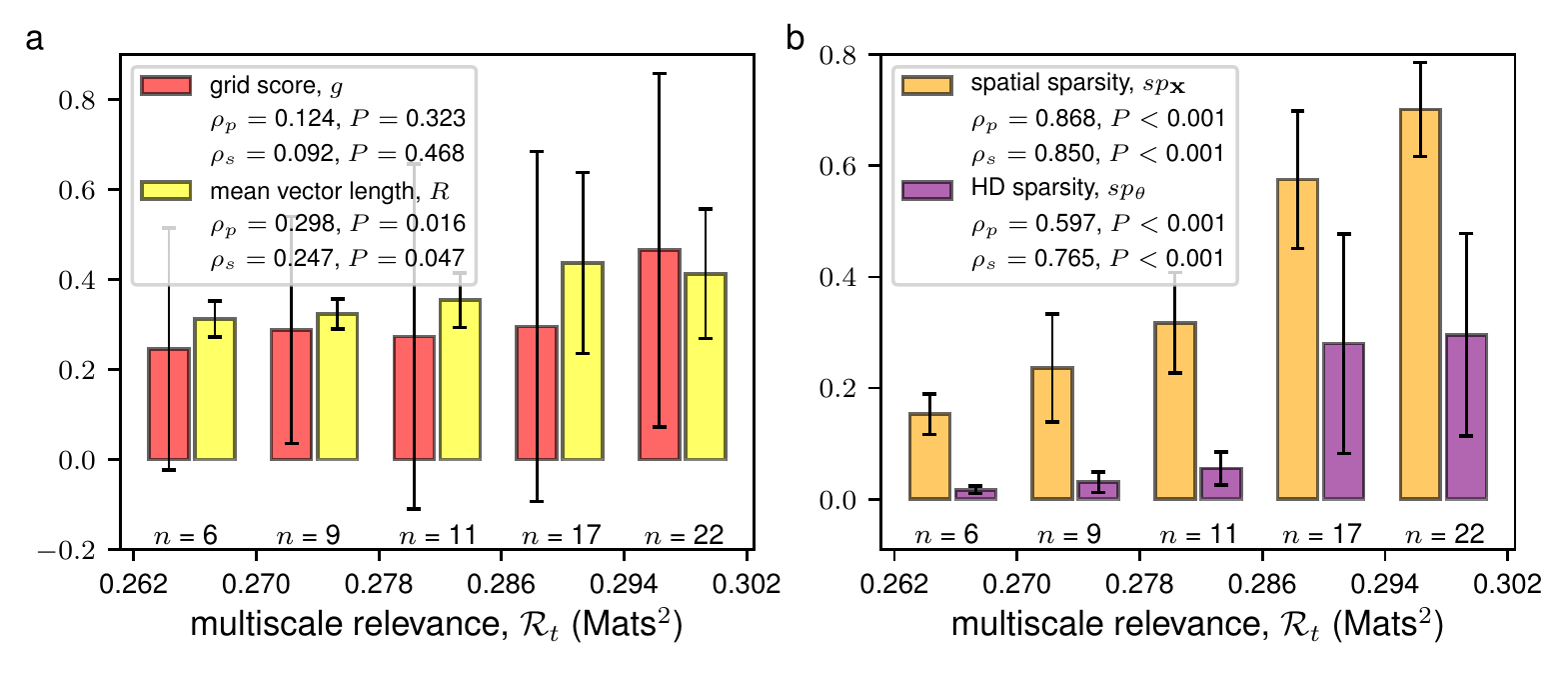}
\end{center}
\caption{\label{figure Figure3}
{\bf The MSR identified neurons with spatially and head directionally selective discharges}.
Bar plots depict the mean (height of the bar) along with the standard deviation (black error bars) of the grid score (red) and Rayleigh mean vector length (yellow) in panel {\bf a}, and the spatial sparsity (orange) and HD sparsity (purple) in panel {\bf b} for each neuron in the mEC within the relevance range as indicated.
The relevance range was determined by equally dividing the range of the calculated MSR into 5 equal parts.
The number of neurons whose MSRs fall within a relevance range is indicated below each bar.
The linearity and monotonicity between the MSR and the different spatial and HD quantities were quantified using the Pearson's correlation, $\rho_p$, and the Spearman's correlation, $\rho_s$, respectively.
}
\end{figure*}

Many of the grid cells were spotted as RNs, but not all.
For example, grid cells 41, 42 and 61, that had a significant grid score, had a low MSR and a low spatial information.
This indicated that different measures correlate differently with MSR.
Fig. \ref{figure Figure3} reports the distribution of the other four measures analyzed in this study conditional to different levels of MSR.
Fig. \ref{figure Figure3}a shows that grid score maintains a large variation across all scales of the MSR, with a moderate increase in its average.
A similar behavior was observed in Fig. \ref{figure Figure3}a for the mean vector length. 
The converse is also true. For example, grid cells 33 and 41 have the same grid score but very different value of the MSR and of the spatial and HD information. A closer inspection of their rate maps (see Fig. S4) substantiates these differences\footnote{Grid cell 41 has a higher firing rate, but also the residual MSR wrt to $\log M$ is significantly negative.}. The $H[s]-H[K]$ curve for neuron 33 stays above the one for neuron 41 at all values of $H[s]$ (see Fig. \ref{figure FigureS9}a). High MSR neurons have very similar $H[s]-H[K]$ curves, which saturate maximal achievable $H[K]$ \citep*{cubero2019criticality}, whereas low MSR neurons differ in characteristic ways. In particular, most of the interneurons feature the same linear $H[s]-H[K]$ relation over an extended range of $H[s]$ shown in Fig. \ref{figure Figure1}c for interneuron 8 (see Fig. \ref{figure FigureS9}d).

Spatial sparsity and HD sparsity, instead, exhibit a significant correlation with the MSR as seen in Fig. \ref{figure Figure3}b. 
The observation that RNs with highly sparse firing may have either low mean vector lengths or low grid scores was an indication that a non-trivial variabilities in firing behaviors does not necessarily obey the imposed symmetries of the tuning curves.

\begin{figure*}[!h]
\begin{center}
\includegraphics[width=\textwidth]{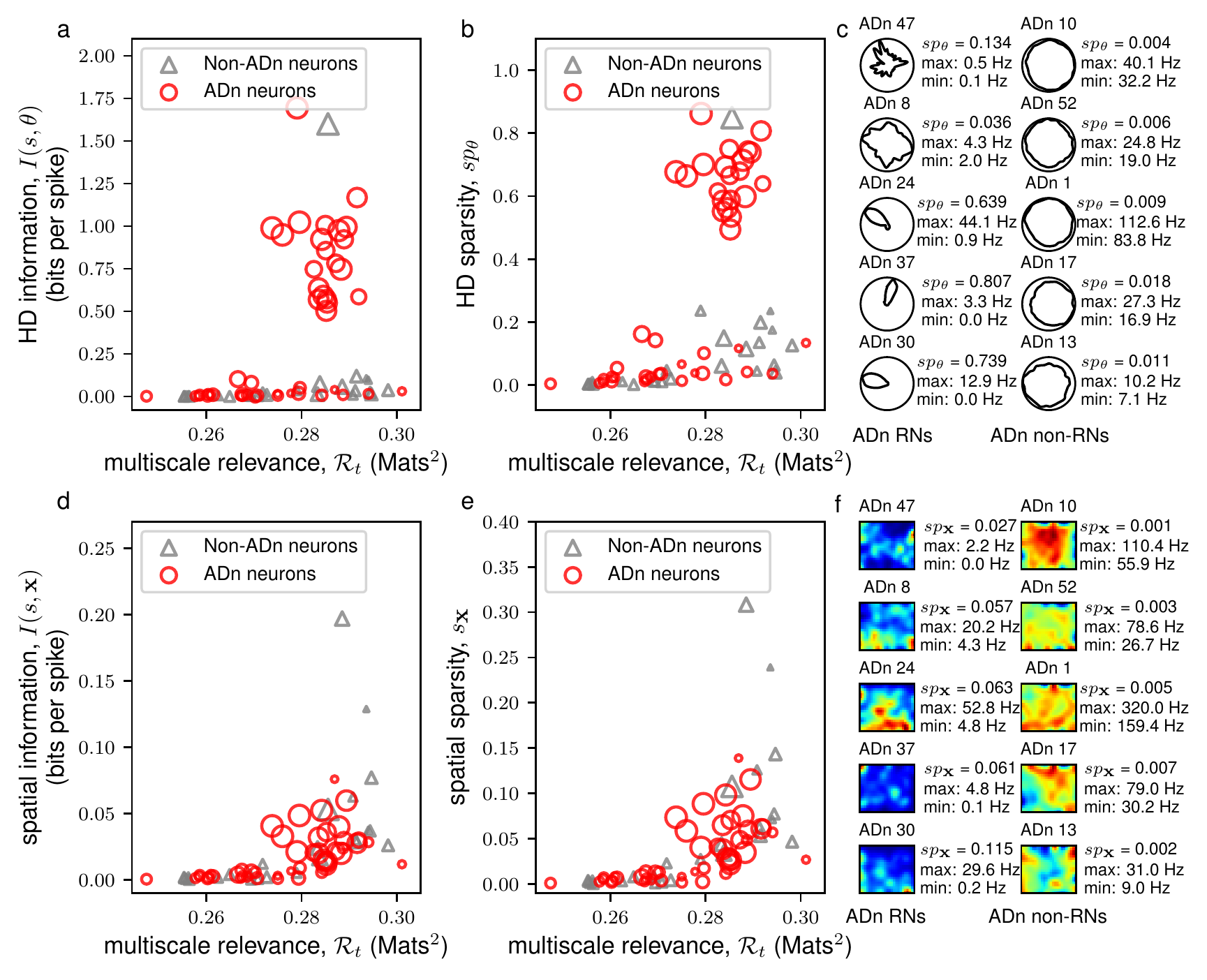}
\end{center}
\caption{\label{figure Figure4}
{\bf MSR of neurons from the anterodorsal thalamic nucleus (ADn) of Mouse 12 from a single recording session (Session 120806)}.
A scatter plot of the multiscale relevance vs. the HD (spatial) information is shown in {\bf a} ({\bf d}).
This plot is supplemented by a scatter plot between the MSR and HD (spatial) sparsity shown in {\bf b} ({\bf e}).
The sizes of the scatter points reflect the mean vector length of the neural activity where the larger scatter points correspond to a sharp preferential firing to a single direction.
The HD tuning curves (spatial firing rate maps) of the 5 most relevant neurons (RNs) and the 5 most irrelevant neurons (non-RNs) are shown together in panel {\bf c} ({\bf f}) together with the calculated HD sparsity, $sp_\theta$, (spatial sparsity, $sp_{\mathrm{\mathbf x}}$) and maximum and minimum firing.
}
\end{figure*}

\begin{figure*}[!h]
\begin{center}
\includegraphics[width=\textwidth]{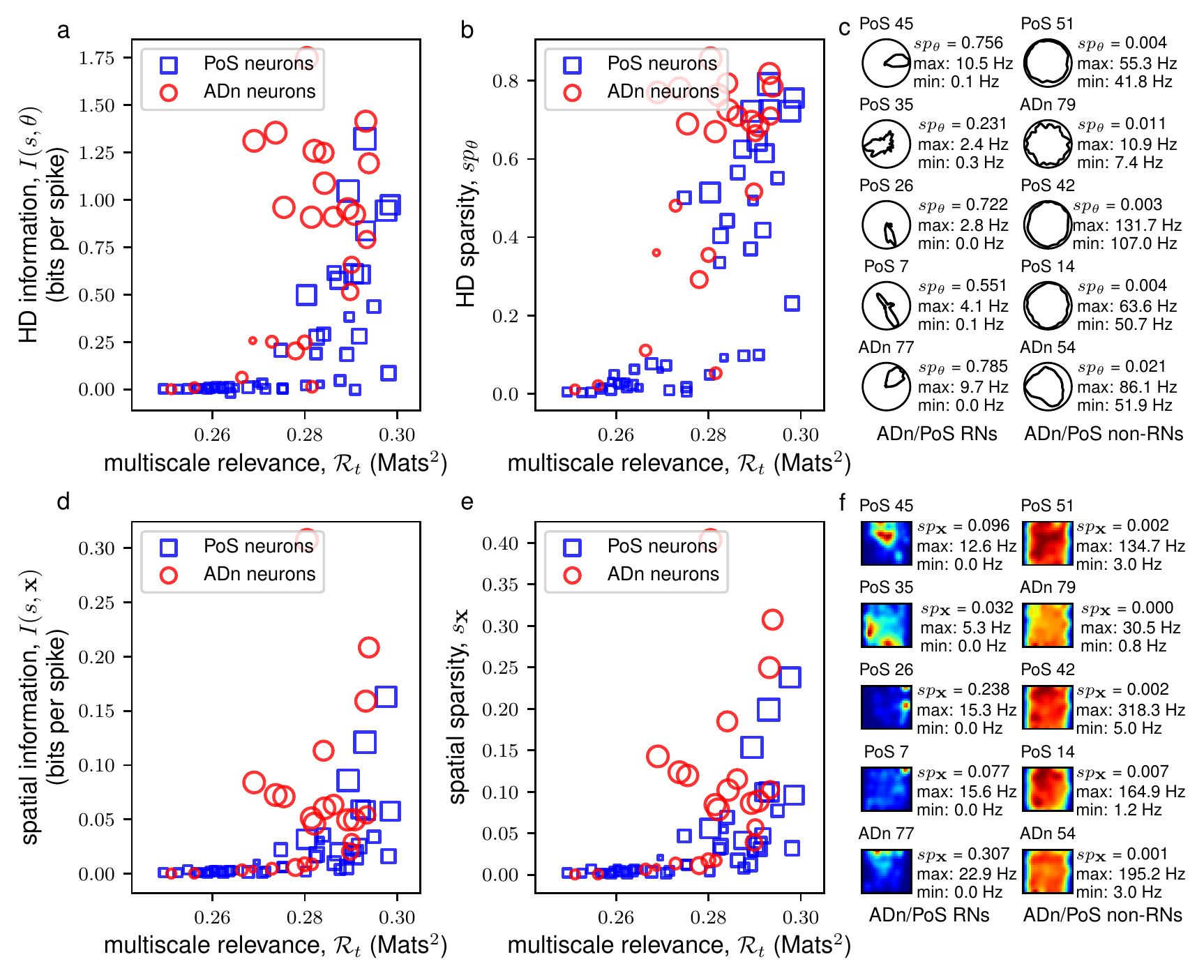}
\end{center}
\caption{\label{figure Figure5}
{\bf MSR of neurons from the anterodorsal thalamic nucleus (ADn) and post-subiculum (PoS) of Mouse 28 from a single recording session (Session 140313)}.
A scatter plot of the MSR vs. the HD (spatial) information is shown in {\bf a} ({\bf d}).
This plot is supplemented by a scatter plot between the MSR and HD (spatial) sparsity shown in {\bf b} ({\bf e}).
The sizes of the scatter points reflect the mean vector length of the neural activity where the larger scatter points correspond to putative head direction cells while the shapes of the scatter points indicate the region where the neuron units were recorded from \citet*{peyrache2015internally} and \citet*{peyrache2015th1}.
The HD tuning curves (spatial firing rate maps) of the 5 most relevant neurons (RNs) and the 5 most irrelevant neurons (non-RNs) are shown together in panel {\bf c} ({\bf f}) together with the calculated HD sparsity, $sp_\theta$, (spatial sparsity, $sp_{\mathrm{\mathbf x}}$) and maximum and minimum firing.
}
\end{figure*}

Following the observations in the mEC, we turned to other regions in the brain -- the thalamus -- to check whether the non-trivial variability revealed by the MSR in the neural spiking, captured functionally relevant external correlates.
To this, we analyzed the neurons in the ADn and PoS areas of freely behaving and navigating rodents.
These regions are known to contain cells that robustly fire when the animal's head is facing a specific direction \citep*{taube1990head,taube1995head} and is believed to be crucial to the formation of grid cells in the mEC \citep*{sargolini2006conjunctive,langston2010development,mcnaughton2006path}.
Thus, we sought to find whether the variability as measured by the MSR contains signals of HD tuning.
We observed that, in all of the 6 mice that were analyzed, the neurons having HD specific firing, i.e., neurons having high HD sparsity and high mean vector lengths, were RNs (see Fig. \ref{figure FigureS6}).
Focusing on a subset of neurons of Mouse 12 (in Fig. \ref{figure FigureS6}a) that were simultaneously recorded in a single session (Session 120806), we observed, as in Figs. \ref{figure Figure4}a,b, that HD attuned neurons were RNs.
However, the HD alone may not explain the structure of the spike frequencies of these neurons \citep*{peyrache2017transformation}.
Hence, we also sought to find whether some of these neurons are spatially tuned.
As seen in Figs. \ref{figure Figure4}d,e, we found that some of the RNs were also modulated by the spatial location of the mouse.
These results were also consistent for a subset of neurons of Mouse 28 (in Fig. \ref{figure FigureS6}f) that were simultaneously recorded in a single session (Session 140313) as in Fig. \ref{figure Figure5}.

\begin{figure*}[!h]
\begin{center}
\includegraphics[width=\textwidth]{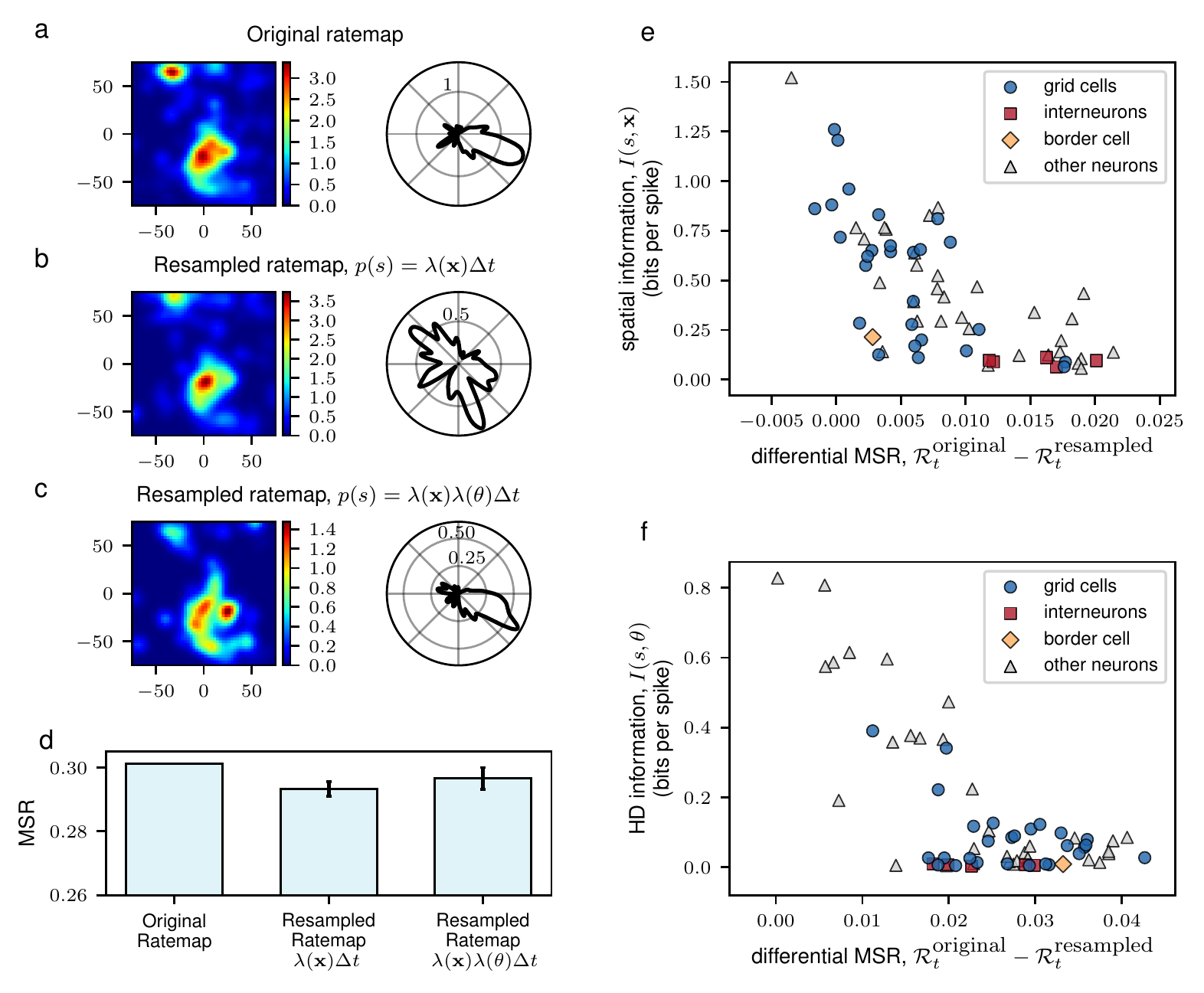}
\end{center}
\caption{\label{figure Figure6}
{\bf The MSR is a measure of information content of the neural activity}.
Resampling the firing rate map using spatial position only or in combination with HD resulted to a firing activity that closely resembled the actual firing pattern of mEC Neuron 47.
Compared to the original firing rate maps in {\bf a}, the spatial (left panels) and HD (right panels) firing rate maps were recovered by the resampling procedure in {\bf b} and {\bf c}.
The result for a single realization of the resampling procedure is shown.
({\bf d}) Bar plots show the MSR calculated from the original spiking activity of the neuron and the resampled rate maps.
The mean and standard deviation of 100 realizations of the resampling procedure is reported.
Scatter plot between the difference of the MSRs of the original spikes and of the synthetic spikes, resampled using only positional information (only HD information), for each neuron and the spatial information is shown in {\bf e} ({\bf f}).
}
\end{figure*}

To assess whether the variations in the spike frequencies, as characterized by the MSR, contained information about external stimuli relevant to navigation, we generated synthetic time series from idealized HD cells and found that neurons with a sharper HD tuning curves have both higher mutual information and higher MSR (see Fig. \ref{figure FigureS1}g). Following this observation, we resampled the spike count code of the neurons in the mEC such that only spatial information, or only HD information, or both spatial and HD information were incorporated.
This resampling of the neural spiking was done by generating synthetic spikes assuming a non-homogeneous Poisson spiking with rates taken from the computed spatial firing rate maps and HD tuning curves (see Section \ref{methods resampling}).
These assumptions were able to approximately recover the original rate maps as seen in Figs. \ref{figure Figure6}b and c.
Here, we focused our attention on \emph{mEC Neuron 47} in the mEC data which had the highest MSR and also had both high spatial and high HD information. However, the same observation can be applied to other neurons that had both high spatial and HD information.

By resampling solely the spatial firing rate map as in Fig. \ref{figure Figure6}d, we saw a decrease in the MSR despite having as much spatial information as the original code.
When HD information was incorporated into the resampled spike frequencies, assuming the factorization of the firing probabilities due to position and HD, we observed an increase in the MSR for \emph{Neuron 47}, almost up to the MSR for the original code. Such increase reveals the additional structure added onto the spiking activity of the resampled neuron.
These findings support the idea that the temporal structure of the spike counts of the neuron, as measured by the MSR, come from its tuning profiles for both position and HD.

We also assessed which cells among the neurons in the mEC have MSRs that could be explained well by the spatial information and thus, were highly spatially attuned.
We resampled the spatial firing rate maps of each of the neurons in the mEC data (see Section \ref{methods resampling}).
The absolute difference between the original and resampled MSR, $ \mathcal{R}_t^{\rm original} - \mathcal{R}_t^{\rm resampled} $, was then computed from the resampled spikes.
When the variations in the spike frequencies could be explained by the spatial firing fields, we expected this difference to be close to zero.
As seen in Fig. \ref{figure Figure6}e, we found that neurons having either high spatial (Fig. \ref{figure Figure6}e) or HD (Fig. \ref{figure Figure6}f) information tended to have a value of the differential MSR $ \mathcal{R}_t^{\rm original} - \mathcal{R}_t^{\rm resampled} $ close to zero.
We also observed that most of the neurons having low differential MSRs were grid cells.
The same observations could be drawn when resampling the HD tuning curves of each of the neurons in the mEC data.
In particular, we also found that neurons having high HD information had differential MSRs close to zero as in Fig. \ref{figure Figure6}f.

Taken altogether, these results suggest that the MSR can be used to identify the interesting neurons in a heterogeneous ensemble.
The proposed measure is able to capture the non-trivial spike frequency distribution across multiple scales whose structure is highly influenced by external correlates that modulate the neural activity.
Indeed, these analyses show that the MSR is able to capture information content of the neural spike code.

\subsection{Relevant neurons decode the external correlates as efficiently as informative neurons.}
We found in the previous section that neurons with low MSR had low spatial or HD information while higher MSR could indicate low or high values of spatial or HD information.
In this section, we show that despite this, high MSR can still be used to select neurons that decode position or HD well.
In other words, although high MSR can imply low spatial or HD information, in terms of population decoding, the highly RNs (selected based on only spike frequencies) performs equally well compared to the highly informative neurons (INs, selected using the knowledge of the external covariate).

To understand whether MSR could identify neurons in mEC whose firing activity allows the animal to identify its position, we compared the decoding efficiency of the 20 neurons with the highest MSR (top RNs) with that of the 20 neurons with the highest spatial information (top spatial INs) wherein the two sets overlap on 14 neurons (see Fig. \ref{figure FigureS7}a).

\begin{figure*}[!h]
\begin{center}
\includegraphics[width=0.9\textwidth]{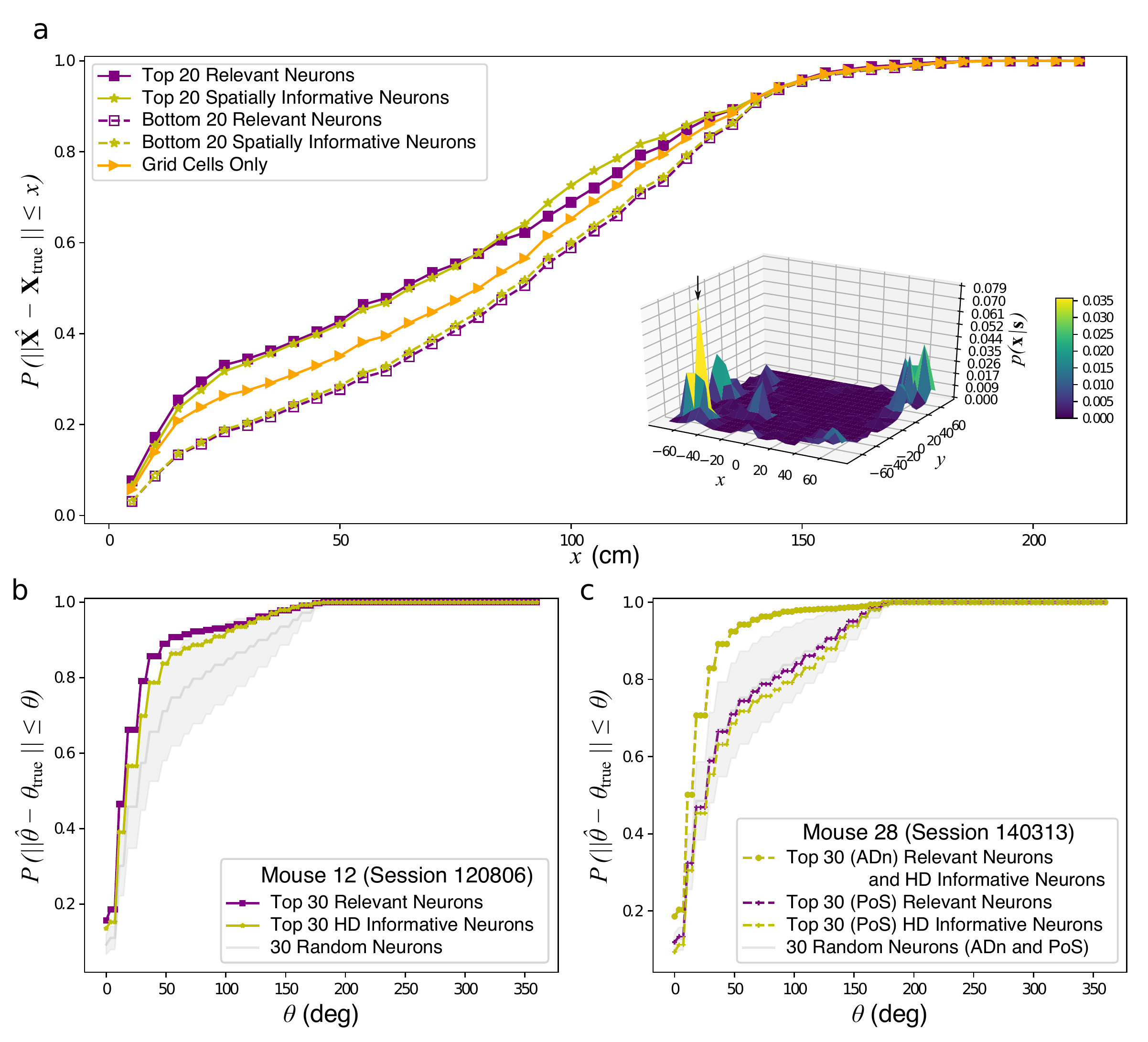}
\end{center}
\caption{\label{figure Figure7}
{\bf Positional decoding of RNs and INs in the mEC and HD decoding of the RNs and INs in the ADn of Mouse 12 and the ADn and PoS of Mouse 28 under a single recording session}.
Panel {\bf a} shows the cumulative distribution of the decoding error, $\| \hat{\mathrm{\mathbf X}} - {\mathrm{\mathbf X}}_{true} \|$, for the RNs (solid violet squares) and spatially INs (solid yellow stars) neurons as well as for the non-RNs (dashed violet squares) and non-INs (dashed yellow stars).
Spatial decoding was also performed for the 27 grid cells in the mEC data (solid orange triangles).
The low positional decoding efficiency at some time points can be traced to the posterior distribution, $p(\mathrm{\mathbf x} \vert \mathrm{\mathbf s})$, of the rat's position given the neural responses which exhibited multiple peaks as shown in the inset surface plot.
For this particular example, the true position was found close to the maximal point of the surface plot as indicated by the arrows although such was not always the case.
Panel {\bf b} depicts the cumulative distribution of the decoding errors of the 30 RNs (violet squares) and 30 HD INs (yellow stars) in the ADn of Mouse 12 in Session 120806.
The mean and standard errors of the cumulative distribution of decoding errors of 30 randomly selected ADn neuron ($n=1000$ realizations) are shown in grey.
On the other hand, panel {\bf c} depicts the cumulative decoding error distribution of the 30 RNs (violet squares) and 30 HD INs in the ADn (yellow crosses) and PoS (yellow circles) of Mouse 28 in Session 140313.
The mean and standard errors of the cumulative distribution of decoding errors of 30 randomly selected ADn or PoS neuron ($n=1000$ realizations) are shown in grey.
As the random selection included neurons from the ADn, which contain a pure head directional information and can decode the positions better than the neurons in the PoS, the decoding errors from the 30 randomly selected neurons were, on average, comparable to that of the  relevant or head directionally informative PoS neurons.
In all the decoding procedures, time points where all the neurons in the ensemble was silent were discarded in the decoding process. 
}
\end{figure*}

To this end, we employed a Bayesian approach to positional decoding wherein the estimated position at the $j$\textsuperscript{th} time bin, $\hat{\mathrm{\mathbf x}}_j$, is determined by the position, ${\mathrm {\mathbf x}}_j$, which maximizes an \emph{a posteriori} distribution, $p( {\mathrm{\mathbf x}}_j \vert {\mathrm{\mathbf s}}_j )$, conditioned on the spike pattern, ${\mathrm{\mathbf s}}_j$, of a neural ensemble within the $j$\textsuperscript{th} time bin i.e.,
\begin{equation}
\hat{\mathrm{\mathbf x}}_j = \arg \max_{{\mathrm{\mathbf x}}_j} p( {\mathrm{\mathbf x}}_j \vert {\mathrm{\mathbf s}}_j ) = \arg \max_{{\mathrm{\mathbf x}}_j} p({\mathrm{\mathbf s}}_j \vert {\mathrm{\mathbf x}}_j )p({\mathrm{\mathbf x}}_j)
\label{map-decoder}
\end{equation}
where the last term is due to Bayes rule, $p({\mathrm{\mathbf s}}_j \vert {\mathrm{\mathbf x}}_j )$ is the likelihood of a spike pattern, ${\mathrm{\mathbf s}}_j$, given the position, ${\mathrm{\mathbf x}}_j$, which depends on a given neuron model and $p({\mathrm{\mathbf x}}_j)$ is the positional occupation probability which can be estimated directly from the data.
Fig. \ref{figure Figure7}a shows that the top RNs decoded just as efficient as the top spatial INs.
It can also be observed that the top RNs decode the positions better than the ensemble composed solely of grid cells.

Because of the sizable overlap between the top RNs and the top spatial INs, one might argue that much of the spatial information needed for positional decoding is concentrated on the neurons in the overlap (ONs).
To address this, we randomly selected 6 neurons among the mEC neurons outside the overlap and, together with the 14 ONs, decoded for the position as done above.
If the positional decoding information is contained in the ONs, then we should observe the same decoding efficiency as either the top RNs or top spatial INs.
However, we found that the decoding efficiency of the ONs decreased (see Fig. \ref{figure FigureS7}d).
We also found that for the decoded positions within 5 cm from the true position, the decoding efficiency of the top RNs were up to 4$\sigma$ from the mean decoding efficiency of the ONs, as measured by the $z$-score compared to that of the top spatial INs which was at around 2$\sigma$.
This indicates that the 6 RNs outside the overlap provide better decodable spatial representation than those of the 6 spatial INs.

Because local variations in the interspike intervals of the neurons in the mEC correlated with spatial information, we sought to find whether neurons with high local variations (LVNs) also contained decodable spatial representation.
We took the top 20 LVNs in the mEC and decoded for the position as done above.
We found that the decoding efficiency of top LVNs are much lower compared to top RNs (see Fig. \ref{figure FigureS3}c).
This indicates that the repertoire of responses coming from local variations in the interspike intervals of mEC neurons alone can not represent space in freely-behaving rats.

To substantiate the decoding results obtained for neurons in the mEC, we also took the ADn RNs and the HD INs in the of Mouse 12 (Session 120806) in Fig. \ref{figure Figure4} to decode for HD.
Mouse 12 was chosen as this animal had the most HD cells recorded among the mice that only had recordings in the ADn \citep*{peyrache2015th1}.
In particular, we looked at the HD decoding at longer time scales (in this case, $\Delta t=100$ ms), where we could model the neural activity using a Poisson distribution, $p({\mathrm{\mathbf n}}_j \vert \theta_j )$ similar to that in Eq. \eqref{decoding-poisson}. Bayesian decoding adopts an equation 
\begin{equation}
\hat{\theta}_j = \arg \max_{\theta_j} p( \theta_j \vert {\mathrm{\mathbf n}}_j ) = \arg \max_{\theta_j} p({\mathrm{\mathbf n}}_j \vert \theta_j )p(\theta_j)
\end{equation}
similar to Eq. \eqref{map-decoder} to estimate the decoded HD, $\hat{\theta}_j$, where $p(\theta_j)$ is the HD occupation as estimated from the data.
We compared the decoding efficiency of the 30 RNs with the 30 HD INs which had 22 neurons that are relevant (see Fig. \ref{figure FigureS3}b).
We also compared the decoding efficiencies of the ADn RNs or HD INs with 30 randomly selected ADn neurons ($n=1000$ realizations).
As seen in Fig. \ref{figure Figure7}b, the RNs decoded just as well as the neural population composed of HD INs.
Furthermore, the decoding efficiency of the RNs were observed to be far better than the decoding efficiency of a random selection of neurons in the ensemble.

We also compared the decoding efficiency of the ADn and PoS neurons from Mouse 28 (Session 140313) which had the most HD cells recorded among the mice that had recordings in both ADn and PoS \citep*{peyrache2015th1} as in Fig. \ref{figure Figure5}.
As seen in Fig. \ref{figure Figure7}c, neurons in the ADn decoded the HD more efficiently than the neurons in the PoS.
These results are consistent with the notion that the ADn contains pure HD modulation which allow for neurons in the ADn to better predict the mouse’s HD compared to the neurons in the PoS which contain, instead, true spatial information \citep*{peyrache2015internally,peyrache2017transformation}.
For the neurons in Mouse 28 (Session 140313), it had to be noted that the 30 ADn RNs also happened to be the 30 ADn HD INs (see Fig. \ref{figure FigureS7}c).
On the other hand, among the 30 PoS RNs, 23 were HD INs (also see Fig. \ref{figure FigureS7}c).
We observed that the PoS RNs decode just as efficient as the PoS HD INs consistent with the findings for Mouse 12 (Session 120806).

Taken altogether, despite being blind to the rat's position and of the mouse's HD, the MSR is able to capture neurons that can decode the position and HD just as well as the spatial INs and as the HD INs.

\section{Discussion}
Hafez wrote the Divan in a manner that the interpretation is left purely to the reader. In this present work, we take inspiration from the Divan by analysing a complex data in a featureless manner such that the data is allowed to speak for what is important in it. In particular, we introduced a novel, parameter-free and fully featureless method -- which we called multiscale relevance (MSR) -- to characterize the temporal structure of the activities of neurons within a heterogeneous population.
We have shown that the neurons showing persistently broad spike frequency distributions across a wide range of time scales, as measured by the MSR, typically carry information about the external correlates related to the behavior of the observed animal. 
By analyzing the neurons in the mEC and nearby brain regions and the neurons in the ADn and PoS -- areas in the brain that are pertinent to spatial navigation -- we showed that the RNs in these regions have firing behaviors that are selective for spatial location and HD.
Here, we found that in many cases, the neurons that display broad spike distributions tend to have conjugated representations in that they exhibit high mutual information with multiple behavioral features.
These findings are consistent with those observed experimentally by \citet*{sargolini2006conjunctive} and statistically by \citet*{hardcastle2017multiplexed}.

The fact that the MSR can be used to select informative neurons as well as neurons that show high decoding performance is consistent with the expectation that the information carried by the activity of a given neuron is encapsulated in the sole spike activity -- the only information available to downstream neurons -- to decode a representation of the feature space. This suggests that relevant neurons should feature a rich variety of long-ranged statistical patterns of the spike activity. This, in turn, results in broad frequency distributions at different time-scales, which are quantified by the relevance $H[K]$, as discussed in \citet*{cubero2019criticality}.
%, we used the ideas of \citet*{cubero2019criticality} and \citet*{marsili2013sampling} to hypothesize that neurons having such non-trivial temporal structures, as manifested by broad distributions of the neural firing behavior, are important to the representations that the brain region encodes.
Hence, at a given resolution, as defined in Eq. (\ref{resolution1}), we estimate the complexity of the temporal code by the relevance defined in Eq. (\ref{relevance}).
%The latter captures the broadness of the spike frequency distribution at that resolution.
Since natural and dynamic stimuli and behaviors often operate on multiple time scales, the MSR integrates over different resolution scales, thus allowing us to spot neurons exhibiting persistent non-trivial spike codes across a broad range of time scales.

Broad distributions of spike frequencies, characterized by a high MSR, exhibit a stochastic variablility that requires richer parametric models \citep*{haimovici2015criticality}.
In a decoding perspective, these non-trivial distributions afford a higher degree of distinguishability of neural responses to a given stimuli or behavior.
Indeed, by decoding either spatial position or HD using statistical approaches, we found that the responses of the RNs allow downstream processing units to efficiently decode the external correlates just as well as the neurons whose resulting tuning maps contain information about those external correlates.

Finally, we observed that the population of relevant neurons, as identified by the MSR, is not homogeneous, e.g., the relevant neurons in the mEC data are not composed solely by grid cells and the relevant neurons in the ADn and PoS are not necessarily composed solely of HD cells.
Noteworthy, the decoding efficiency of the relevant neurons was observed to be better compared to the ensemble comprising solely the grid cells.
When taken altogether, these observations support the idea that population heterogeneity may play a role towards efficient encoding of stimuli \citep*{chelaru2008efficient,meshulam2017collective}.

The fact that the MSR captures functional information from the temporal code is a remarkable aspect of this measure.
This method can then be used as a pre-processing tool to impose a less stringent criteria compared to those widely used in many studies (e.g., mean vector length, spatial sparsity and grid scores) thereby directing further investigation to interesting neurons.
The MSR is expected to be particularly useful in detecting relevant neurons in high-throughput studies -- where the activity of many neurons are measured simultaneously or in single-electrode neural recordings where, under a given task, an experiment is done multiple times -- where the function of neurons or the correlates they encode are not  known {\em a priori}. Our discussion has been confined to correlates related to navigation, but it applies in a straightforward manner to other correlates (e.g. heart rate or pupil diameter), that may be responsible for the recorded activity of relevant neurons. 

%Furthermore, while the current application focuses on multiple-electrode experiments, we also expect MSR to be useful even in single-electrode neural recordings where, under a given task, an experiment is done multiple times. Since the variations as captured by the MSR are a possible signature of a neuron under a particular task or behavior, the MSR can be used to assess whether the electrode was struck on the right brain region or not.

Whether this measure can also be used to identify functionally relevant neuronal units recorded through calcium imaging or through fMRI is also an exciting direction for future studies. A further promising direction lies in the extension of the principles used to construct MSR to the study of neural assemblies \citep*{russo2017cell}. This could allow one to probe the importance of correlated firing of neurons in representing external stimuli or behaviors. For example, the analysis of boolean functions of pairs of neurons in the mEC recordings, shown in the Supplementary Material (Fig. \ref{figure FigureS8}), suggests that the representations of individual neurons are non-redundant and that interneurons play a peculiar role in the information aggregation. 

\section{Materials and methods}
\subsection{Data Collection}
\label{methods collection}
The data used in this study are recordings from rodents with multisite tetrode implants.
These neurons are of particular interest because they are involved in spatial navigation.

\subsubsection{Data from medial entorhinal cortex (mEC)}
The spike times of 65 neurons recorded across the mEC area of a male Long Evans rat (Rat 14147) were taken from \citet*{stensola2012entorhinal}.
The rat was allowed to freely explore a box of dimension $150~\times~150$ cm\textsuperscript{2} for a duration of around 20 mins.
The positions were tracked using a platform attached to the head with red and green diodes fixed at both ends.
Additional details about the data acquisition can be found in the paper by \citet*{stensola2012entorhinal}.

\subsubsection{Data from the anterodorsal thalamic nucleus (ADn) and post-subiculum (PoS)}
The spike times of 746 neurons recorded from multiple areas in the ADn and PoS across multiple sessions in six free moving mice (Mouse 12, Mouse 17, Mouse 20, Mouse 24, Mouse 25 and Mouse 28) while they freely foraged for food across an open environment with dimensions $53~\times~46$ cm\textsuperscript{2} and in their home cages during sleep were taken from \citet*{peyrache2015th1}.
Mouse 12, Mouse 17 and Mouse 20 only had recordings in the ADn while Mouse 24, Mouse 25 and Mouse 28 had simultaneous recordings from ADn and PoS.
The positions were tracked using a platform attached to the heads of the mice with red and blue diodes fixed at both ends.
Only the recorded spike times during awake sessions and the neural units with at least 100 observed spikes were considered in this study.
Additional information regarding the data acquisition can be found in the paper by \citet*{peyrache2015internally} and the CRCNS\footnote{https://crcns.org} database entry by \citet*{peyrache2015th1}.

\subsection{Position and speed filtering}
\label{methods filtering}
The position time series for the mEC data were smoothed to reduce jitter using a low-pass Hann window FIR filter with cutoff frequency of 2.0 Hz and kernel support of 13 taps (approximately 0.5 s) and were then renormalized to fill missing bins within the kernel duration as done by \citet*{dunn2017grid}.
The rat's position was taken to be the average of the recorded and filtered positions of the two tracked diodes.
The head direction was calculated as the angle of the perpendicular bisector of the line connecting the two diodes using the filtered positions.
The speed at each time point was computed by dividing the trajectory length with the elapsed time within a 13-time point window.
When calculating for spatial firing rate maps and spatial information (see below), only time points where the rat was running faster than 5 cm/s were considered.
No speed filters were imposed when calculating for head directional tuning curves and head directional information.
On the other hand, no position smoothing nor speed filtering were performed when calculating for the spatial firing rate maps and spatial information for the ADn and PoS data.

\subsection{Rate maps}
\label{methods ratemaps}
The spike location, $\mathbf{\xi}_j^{(i)}$, of neuron $i$ at a spike time $t_j^{(i)}$ was calculated by linearly interpolating the filtered position time series at the spike time.
As done by \citet*{dunn2017grid}, the spatial firing rate map at position $\mathrm{\mathbf x} = (x,y)$ was calculated as the ratio of the kernel density estimates of the spatial spike frequency and the spatial occupancy, both binned using $3$ cm square bins, as
\begin{equation}
f(\mathrm{\mathbf x}) = \frac{\sum_{j=1}^{M} K( \mathrm{\mathbf x} \vert \mathbf{\xi}_j )}{\sum_{j=1}^{M} \Delta t_j K({\mathrm{\mathbf x}} \vert {\mathrm{\mathbf x}}_j )}
\end{equation}
where a triweight kernel
\begin{equation}
K( \mathrm{\mathbf x} \vert \mathbf{\xi} ) = \frac{4}{9\pi\sigma_K^2} \left[ 1 - \frac{ \| \mathrm{\mathbf x} - \mathbf{\xi} \|^2 }{9\sigma_K^2}\right]^3, \| \mathrm{\mathbf x} - \mathbf{\xi} \| < 3\sigma_K 
\end{equation}
with bandwidth $\sigma_K = 4.2$ cm was used.
In place of a triweight kernel, a Gaussian smoothing kernel with $\sigma_G = 4.0$ truncated at $4\sigma_G$ was also used to estimate the rate maps which gave qualitatively similar results.
For better visualization, a Gaussian smoothing kernel with $\sigma_G = 8.0$ was used to filter the spatial firing rate map.

On the other hand, for head direction tuning curves, the angles were binned using $9^{\circ}$ bins.
The tuning curve was then calculated as the ratio of the head direction spike frequency and the head direction occupancy without any smoothing kernels as the head direction bins are sampled well-enough.
For better visualization, a Gaussian  kernel with smoothing window of $20^{\circ}$ was used to filter the tuning curves.

\subsection{Information, Sparsity and other Scores}
\label{methods scores}
Given a feature, $\phi$ (e.g., spatial position, $\mathrm{\mathbf x}$, head direction, $\theta$ or speed, $v$), the information between the neural spiking $\mathrm{\mathbf s}$ and the feature can be calculated \'a la Skaggs-McNaughton \citep*{skaggs1993information}.
In particular, under the assumption of a non-homogeneous Poisson process with feature dependent rates, $\lambda(\phi)$, under small time intervals $\Delta t$, the amount of information, in bits per second, that can be decoded from the rate maps is given by
\begin{equation}
I(s, \phi) = \sum_{\phi} p(\phi)\frac{\lambda(\phi)}{\bar{\lambda}}  \log{\frac{\lambda(\phi)}{\bar{\lambda}}}
\label{equation information}
\end{equation}
where $\lambda(\phi)$ is the firing rate at $\phi$, $p(\phi)$ is the probability of occupying $\phi$ and 
\begin{equation}
\bar{\lambda} \equiv \sum_\phi \lambda(\phi) p(\phi)
\label{equation firing rate}
\end{equation}
is the average firing rate.
To account for the bias due to finite samples, the information of a randomized spike frequency was calculated using a bootstrapping procedure.
To this end, the spikes were randomly shuffled 1000 times and the information for each reshuffling was calculated.
The average randomized information was then subtracted from the non-randomized information. It is interesting to mention, in passing, that reshuffling wipes all information between all correlates and the time of spiking. Since the MSR only depends on the timing of the spikes, and not on other correlates, is it unaffected by reshuffling. 

Apart from the information, one of the measures that are used to quantify selectivity of neural firing to a given feature is the firing sparsity \citep*{buetfering2014parvalbumin} which can be calculated using
\begin{equation}
sp_\phi = 1 - \frac{ \left( \sum_\phi \lambda(\phi) p(\phi) \right)^2}{\sum_\phi \lambda(\phi)^2 p(\phi)}.
\label{equation sparsity}
\end{equation}

Apart from the measures of information and sparsity, we also calculated the grid scores, $g$, for the neurons in the mEC data.
The grid score is designed to quantify the hexagonality of the spatial firing rate maps through the spatial autocorrelation maps (or autocorrelograms) and was first used by \citet*{sargolini2006conjunctive} to identify putative grid cells.
In brief, the grid score is computed from the spatial autocorrelogram where each element $\rho_{ij}$ is the Pearson's correlation of overlapping regions between the spatial firing rate map shifted $i$ bins in the horizontal axis and $j$ bins in the vertical axis and the unshifted rate map.
The angular Pearson autocorrelation, ${\rm acorr}(u)$, of the spatial autocorrelogram was then calculated using spatial bins within a radius $u$ from the center at lags (or rotations) of $30^{\circ}$, $60^{\circ}$, $90^{\circ}$, $120^{\circ}$ and $150^{\circ}$, as well as the $\pm 3^{\circ}$ and $\pm 6^{\circ}$ offsets from these angles to account for sheared grid fields \citep*{stensola2015shearing}.
As done by \citet*{dunn2017grid}, the grid score, $g(u)$, for a fixed radius of $u$, is computed as
\begin{align}
g(u) &= \frac{1}{2} \left[ \max \lbrace \mathrm{acorr}(u) \mathrm{~at~} 60^{\circ} \pm (0^{\circ}, 3^{\circ}, 6^{\circ}) + \right. \nonumber \\
& \qquad \left. \max \lbrace \mathrm{acorr}(u) \mathrm{~at~} 120^{\circ} \pm (0^{\circ}, 3^{\circ}, 6^{\circ}) \right] \nonumber \\
&- \frac{1}{3} \left[ \min \lbrace \mathrm{acorr}(u) \mathrm{~at~} 30^{\circ} \pm (0^{\circ}, 3^{\circ}, 6^{\circ}) + \right. \nonumber \\
&\qquad \left. \min \lbrace \mathrm{acorr}(u) \mathrm{~at~} 90^{\circ} \pm (0^{\circ}, 3^{\circ}, 6^{\circ}) + \right. \nonumber \\
&\qquad \left. \min \lbrace \mathrm{acorr}(u) \mathrm{~at~} 150^{\circ} \pm (0^{\circ}, 3^{\circ}, 6^{\circ}) \right].
\label{equation gridscore}
\end{align}
The final grid score, $g$, is then taken as the maximal grid score, $g(u)$, within the interval $u \in [12 \mathrm{~cm}, 75 \mathrm{~cm}]$ in intervals of 3 cm.

Another quantity that was calculated in this paper is the Rayleigh mean vector length, $R$.
Given the angles $\lbrace \theta_1, \ldots, \theta_M \rbrace$ where a neuronal spike was recorded, the mean vector length can be calculated as
\begin{equation}
R = \sqrt{ \left( \frac{1}{M} \sum_{i=1}^M \cos\theta_i \right)^2 + \left( \frac{1}{M} \sum_{i=1}^M \sin\theta_i \right)^2 }.
\label{equation meanvectorlength}
\end{equation}
Note that for head direction cells where the neuron fires at a specific head direction, the angles will be mostly concentrated along the preferred head direction, $\theta_c$, and hence, $R \approx 1$ whereas for neurons with no preferred direction, $R \approx 0$.

\subsection{Resampling the firing rate map}
\label{methods resampling}
The calculated rate maps and the real animal trajectory were used to resample the neural activity assuming non-homogeneous Poisson spiking statistics with rates taken from the rate maps.
To this end, the real trajectory of the rat was divided into $\Delta t = 1$ ms bins.
The position and head direction were linearly interpolated from the filtered positions described above.
The target firing rate, $f_j$ in bin $j$ was then calculated by evaluating the tuning profile at the interpolated position or head direction.
Whenever the target firing rate was modulated by both the position and head direction, we assumed that the contribution due to each feature was multiplicative and thus, $f_j$ is calculated as the product of the tuning profiles at the interpolated position and the interpolated head direction.
A Bernoulli trial was then performed in each bin with a success probability given by $f_j \Delta t$.

\subsection{Statistical decoding}
\label{methods decoding}
For positional decoding, we divided the space in a grid of $20\times 20$ cells of $7.5$ cm $\times~ 7.5$ cm spatial resolution, which was comparable to the rat's body length.
Time was also discretized into 20 ms bins which ensured that for most of the time (i.e. in $92\%$ of the cases), the rat was located within a single spatial cell.
Under these time scales, the responses of a neuron can be regarded as being drawn from a binomial distribution, i.e., either the neuron $i$ is active ($s_j^{(i)}=1$) or not ($s_j^{(i)}=0$) between $(j-1) \Delta t$ and $j \Delta t$.
The likelihood of the neural responses, ${\rm \textbf{s}}_j = (s_j^{(1)}, \ldots, s_j^{(N)})$ of $N$ independent neurons at a given time conditioned on the position, ${\mathrm{\mathbf x}}_j$ is then given by
\begin{equation}
p({\mathrm{\mathbf s}}_j \vert {\mathrm{\mathbf x}}_j ) = \prod_{i=1}^N (\lambda^{(i)}({\mathrm{\mathbf x}}_j)\Delta t)^{s^{(i)}_j} (1-\lambda^{(i)}({\mathrm{\mathbf x}}_j)\Delta t)^{1-s^{(i)}_j}
\label{decoding-binomial}
\end{equation}
where $\lambda^{(i)}({\mathrm{\mathbf x}}_j)$ is the firing rate of neuron $i$ at ${\mathrm{\mathbf x}}_j$ estimated from its corresponding spatial firing rate map.
Given the prior distribution on the position, $p({\mathrm{\mathbf x}}_j)$, which is estimated from the data, the posterior distribution of the position, ${\mathrm{\mathbf x}}_j$, given the neural responses, ${\mathrm{\mathbf s}}_j$ at time $t$ is given by
\begin{equation}
p( {\mathrm{\mathbf x}}_j \vert {\mathrm{\mathbf s}}_j ) = \frac{p({\mathrm{\mathbf s}}_j \vert {\mathrm{\mathbf x}}_j )p({\mathrm{\mathbf x}}_j)}{p({\mathrm{\mathbf s}}_j)}.
\end{equation}
The decoded position, as in the Bayesian 1-step decoding by \citet*{zhang1998interpreting}, was calculated as
\begin{equation}
\hat{\mathrm{\mathbf x}}_j = \arg \max_{{\mathrm{\mathbf x}}_j} p({\mathrm{\mathbf s}}_j \vert {\mathrm{\mathbf x}}_j )p({\mathrm{\mathbf x}}_j).
\end{equation}

For head directional decoding, on the other hand, we divided the angles, $\theta \in [0, 2 \pi)$ in $9^{\circ}$ bins.
For this case, time was instead discretized into 100 ms bins. Under these time scales, the neurons could not be regarded simply as either active or not.
Hence, it was natural to switch towards the analysis of population vectors, ${\mathrm{\mathbf n}}_j$, a vector which represents the number of spikes, $n^{(i)}_j$, recorded from each neuron within the $j$\textsuperscript{th} time bin, to decode for the head direction. 
In this case, the number of spikes, $n^{(i)}_j$, that neuron $i$ discharges between $(j-1) \Delta t$ and $j \Delta t$ can be modeled as a non-homogeneous Poisson distribution
\begin{equation}
p(n^{(i)}_j \vert \theta_j ) = \frac{\lambda^{(i)}(\theta_j)^{n^{(i)}_j}}{n^{(i)}_j!} \exp(-\lambda^{(i)}(\theta_j))
\label{decoding-poisson}
\end{equation}
with $\lambda^{(i)}(\theta_j)$ being the firing rate of neuron $i$ at $\theta_j$ estimated from the HD tuning curve, and thus, under the independent neuron assumption, $p({\mathrm{\mathbf n}}_j \vert \theta_j ) = \prod_{i=1}^N p(n^{(i)}_j \vert \theta_j )$.
The decoded head direction can then be calculated as
\begin{equation}
\hat{\theta}_j = \arg \max_{\theta_j} p({\mathrm{\mathbf n}}_j \vert \theta_j )p(\theta_j).
\end{equation}
where $p(\theta_j)$ is the head directional prior distribution which is estimated from the data.
Note that in all of the decoding procedures, we only decoded for time points with which at least one neuron was active. Furthermore, the decoding exercise for both space and HD were done on different time bins, spanning from 10 ms to 200 ms, and obtained qualitatively similar results.

\subsection{Boolean function}
Spikes from mEC neurons were binned and binarized at $\Delta t = 1$ms time intervals such that each bin $B_s$ takes a value of 1 if there is at least one recorded spike within the time interval $[(s-1)\Delta t, s\Delta t)$ and 0 otherwise. Given two neuron pairs $i$ and $j$, a new spike train was built using the Boolean function rules in Table \ref{tab:1} to each time bin. The corresponding MSR, and spatial and HD information were then calculated. The same procedure was done using $\Delta t= 0.5$ ms time intervals which gave qualitatively similar results. Note that the events where two neurons fire together (that are those identified by the AND function) are generally very sparse and do not allow for a reliable calculation of the MSR and thus, the MSR for pairs of neurons having at least 100 co-firing events were considered.

\begin{table}[!h]
% table caption is above the table
\caption{Boolean function rules}
\label{tab:1}       % Give a unique label
% For LaTeX tables use
\begin{tabular}{ccccc}
\hline\noalign{\smallskip}
$i$ & $j$ & ${\rm AND}(i,j)$ & ${\rm OR}(i,j)$ & ${\rm XOR}(i,j)$\\
\noalign{\smallskip}\hline\noalign{\smallskip}
0 & 0 & 0 & 0 & 0\\
0 & 1 & 0 & 1 & 1\\
1 & 0 & 0 & 1 & 1\\
1 & 1 & 1 & 1 & 0\\
\noalign{\smallskip}\hline
\end{tabular}
\end{table}

\subsection{Noise correlations}
Spatial noise correlations were calculated as done by \citet*{dunn2015correlations}. In brief, spikes were binned at $\Delta t = 1$ms time intervals and were smoothened by a Gaussian kernel with width of 20ms. The spatial environment was binned into a grid of 7.5 cm square tiles and the trajectories over the spatial bin $\alpha$, defined as the time that the rat enters and leaves the square tile $\alpha$, were noted. For each neuron $i$, a $1 \times k$ vector, ${\mathrm{\mathbf r}}_i^\alpha$, of the mean firing rate over each of the $k$ trajectories was constructed. The spatial noise correlation between neuron pairs $i$ and $j$ were then calculated as
\begin{equation}
C_{ij}(\mathrm{\mathbf x}) = \langle \rho_P({\mathrm{\mathbf r}}_i^\alpha, {\mathrm{\mathbf r}}_j^\alpha) \rangle_{\alpha}
\end{equation}
where $\rho_P({\mathrm{\mathbf x}}, {\mathrm{\mathbf y}})$ is the Pearson's correlation and the averages are taken over the spatial bins $\alpha$.

\subsection{Source codes}
All the calculations in this manuscript were done using personalized scripts written in Python 3.
The source codes for calculating multiscale relevance (which is also compatible with Python 2) and for reproducing the figures in the main text are accessible online\footnote{https://github.com/rcubero/MSR}.

\section*{Acknowledgements}
This research was supported by the Kavli Foundation and the Centre of Excellence scheme of the Research Council of Norway (Centre for Neural Computation).
RJC is currently receiving funding from the European Union's Horizon 2020 research and innovation programme under the Marie Sk\l{}odowska-Curie Grant Agreement No. 754411.
\bibliographystyle{apalike}

\begin{thebibliography}{}

\bibitem[Battistin et~al., 2017]{battistin2017learning}
Battistin, C., Dunn, B., and Roudi, Y. (2017).
\newblock Learning with unknowns: analyzing biological data in the presence of
  hidden variables.
\newblock {\em Current Opinion in Systems Biology}, 1:122--128.

\bibitem[Buetfering et~al., 2014]{buetfering2014parvalbumin}
Buetfering, C., Allen, K., and Monyer, H. (2014).
\newblock Parvalbumin interneurons provide grid cell--driven recurrent
  inhibition in the medial entorhinal cortex.
\newblock {\em Nature neuroscience}, 17(5):710--718.

\bibitem[Chelaru and Dragoi, 2008]{chelaru2008efficient}
Chelaru, M.~I. and Dragoi, V. (2008).
\newblock Efficient coding in heterogeneous neuronal populations.
\newblock {\em Proceedings of the National Academy of Sciences},
  105(42):16344--16349.

\bibitem[Couey et~al., 2013]{couey2013recurrent}
Couey, J.~J., Witoelar, A., Zhang, S.-J., Zheng, K., Ye, J., Dunn, B.,
  Czajkowski, R., Moser, M.-B., Moser, E.~I., Roudi, Y., et~al. (2013).
\newblock Recurrent inhibitory circuitry as a mechanism for grid formation.
\newblock {\em Nature neuroscience}, 16(3):318--324.

\bibitem[Cover and Thomas, 2012]{cover2012elements}
Cover, T.~M. and Thomas, J.~A. (2012).
\newblock {\em Elements of information theory}.
\newblock John Wiley \& Sons.

\bibitem[Cubero et~al., 2019]{cubero2019criticality}
Cubero, R.~J., Jo, J., Marsili, M., Roudi, Y., and Song, J. (2019).
\newblock Statistical criticality arises in most informative representations.
\newblock {\em Journal of Statistical Mechanics: Theory and Experiment},
  2019(6):P063402.

\bibitem[Cubero et~al., 2018]{cubero2018entropy}
Cubero, R.~J., Marsili, M., and Roudi, Y. (2018).
\newblock Minimum description length codes are critical.
\newblock {\em Entropy}, 20(10).

\bibitem[Dunn et~al., 2015]{dunn2015correlations}
Dunn, B., M{\o}rreaunet, M., and Roudi, Y. (2015).
\newblock Correlations and functional connections in a population of grid
  cells.
\newblock {\em PLoS computational biology}, 11(2):e1004052.

\bibitem[Dunn et~al., 2017]{dunn2017grid}
Dunn, B., Wennberg, D., Huang, Z., and Roudi, Y. (2017).
\newblock Grid cells show field-to-field variability and this explains the
  aperiodic response of inhibitory interneurons.
\newblock {\em arXiv preprint arXiv:1701.04893}.

\bibitem[Ebbesen et~al., 2016]{ebbesen2016cell}
Ebbesen, C.~L., Reifenstein, E.~T., Tang, Q., Burgalossi, A., Ray, S.,
  Schreiber, S., Kempter, R., and Brecht, M. (2016).
\newblock Cell type-specific differences in spike timing and spike shape in the
  rat parasubiculum and superficial medial entorhinal cortex.
\newblock {\em Cell reports}, 16(4):1005--1015.

\bibitem[Goh and Barab{\'a}si, 2008]{goh2008burstiness}
Goh, K.-I. and Barab{\'a}si, A.-L. (2008).
\newblock Burstiness and memory in complex systems.
\newblock {\em EPL (Europhysics Letters)}, 81(4):48002.

\bibitem[Grigolon et~al., 2016]{grigolon2016identifying}
Grigolon, S., Franz, S., and Marsili, M. (2016).
\newblock Identifying relevant positions in proteins by critical variable
  selection.
\newblock {\em Molecular BioSystems}, 12(7):2147--2158.

\bibitem[Hafting et~al., 2005]{hafting2005microstructure}
Hafting, T., Fyhn, M., Molden, S., Moser, M.-B., and Moser, E.~I. (2005).
\newblock Microstructure of a spatial map in the entorhinal cortex.
\newblock {\em Nature}, 436(7052):801.

\bibitem[Haimovici and Marsili, 2015]{haimovici2015criticality}
Haimovici, A. and Marsili, M. (2015).
\newblock Criticality of mostly informative samples: a bayesian model selection
  approach.
\newblock {\em Journal of Statistical Mechanics: Theory and Experiment},
  2015(10):P10013.

\bibitem[Hardcastle et~al., 2017]{hardcastle2017multiplexed}
Hardcastle, K., Maheswaranathan, N., Ganguli, S., and Giocomo, L.~M. (2017).
\newblock A multiplexed, heterogeneous, and adaptive code for navigation in
  medial entorhinal cortex.
\newblock {\em Neuron}, 94(2):375--387.

\bibitem[Hubel and Wiesel, 1959]{hubel1959receptive}
Hubel, D.~H. and Wiesel, T.~N. (1959).
\newblock Receptive fields of single neurones in the cat's striate cortex.
\newblock {\em The Journal of physiology}, 148(3):574--591.

\bibitem[Insanally et~al., 2019]{insanally2019spike}
Insanally, M.~N., Carcea, I., Field, R.~E., Rodgers, C.~C., DePasquale, B., Rajan, K., DeWeese, M.~R., Albanna, B.~F., Froemke, R.~C. (2019).
\newblock Spike-timing-dependent ensemble encoding by non-classically responsive cortical neurons.
\newblock {\em eLife}, 8:e42409.

\bibitem[Kropff et~al., 2015]{kropff2015speed}
Kropff, E., Carmichael, J.~E., Moser, M.-B., and Moser, E.~I. (2015).
\newblock Speed cells in the medial entorhinal cortex.
\newblock {\em Nature}, 523(7561):419--424.

\bibitem[Krupic et~al., 2015]{krupic2015grid}
Krupic, J., Bauza, M., Burton, S., Barry, C., and O'Keefe, J. (2015).
\newblock Grid cell symmetry is shaped by environmental geometry.
\newblock {\em Nature}, 518(7538).

\bibitem[Langston et~al., 2010]{langston2010development}
Langston, R.~F., Ainge, J.~A., Couey, J.~J., Canto, C.~B., Bjerknes, T.~L.,
  Witter, M.~P., Moser, E.~I., and Moser, M.-B. (2010).
\newblock Development of the spatial representation system in the rat.
\newblock {\em Science}, 328(5985):1576--1580.

\bibitem[Latuske et~al., 2015]{latuske2015interspike}
Latuske, P., Toader, O., and Allen, K. (2015).
\newblock Interspike intervals reveal functionally distinct cell populations in
  the medial entorhinal cortex.
\newblock {\em Journal of Neuroscience}, 35(31):10963--10976.

\bibitem[Lederberger et~al., 2018]{lederberger2018}
Lederberger, D., Battistin, C., Gardner, R.~J., Roudi, Y., Witter, M., Moser,
  M.~B., and Moser, E.~I. (2018).
\newblock Multiplexed spatial representations in subiculum.
\newblock FENS abstract.

\bibitem[Marsili et~al., 2013]{marsili2013sampling}
Marsili, M., Mastromatteo, I., and Roudi, Y. (2013).
\newblock On sampling and modeling complex systems.
\newblock {\em Journal of Statistical Mechanics: Theory and Experiment},
  2013(09):P09003.

\bibitem[McNaughton et~al., 2006]{mcnaughton2006path}
McNaughton, B.~L., Battaglia, F.~P., Jensen, O., Moser, E.~I., and Moser, M.-B.
  (2006).
\newblock Path integration and the neural basis of the 'cognitive map'.
\newblock {\em Nature Reviews Neuroscience}, 7(8):663--678.

\bibitem[Mease et~al., 2017]{mease2017multiplexed}
Mease, R.~A., Kuner, T., Fairhall, A.~L., and Groh, A. (2017).
\newblock Multiplexed spike coding and adaptation in the thalamus.
\newblock {\em Cell reports}, 19(6):1130--1140.

\bibitem[Merzenich et~al., 1975]{merzenich1975representation}
Merzenich, M.~M., Knight, P.~L., and Roth, G.~L. (1975).
\newblock Representation of cochlea within primary auditory cortex in the cat.
\newblock {\em Journal of neurophysiology}, 38(2):231--249.

\bibitem[Meshulam et~al., 2017]{meshulam2017collective}
Meshulam, L., Gauthier, J.~L., Brody, C.~D., Tank, D.~W., and Bialek, W.
  (2017).
\newblock Collective behavior of place and non-place neurons in the hippocampal
  network.
\newblock {\em Neuron}, 96(5):1178--1191.

\bibitem[O'Keefe and Dostrovsky, 1971]{o1971hippocampus}
O'Keefe, J. and Dostrovsky, J. (1971).
\newblock The hippocampus as a spatial map: Preliminary evidence from unit
  activity in the freely-moving rat.
\newblock {\em Brain research}, 34:171--175.

\bibitem[Panzeri et~al., 2010]{panzeri2010sensory}
Panzeri, S., Brunel, N., Logothetis, N.~K., and Kayser, C. (2010).
\newblock Sensory neural codes using multiplexed temporal scales.
\newblock {\em Trends in neurosciences}, 33(3):111--120.

\bibitem[Pastoll et~al., 2013]{pastoll2013feedback}
Pastoll, H., Solanka, L., van Rossum, M.~C., and Nolan, M.~F. (2013).
\newblock Feedback inhibition enables theta-nested gamma oscillations and grid
  firing fields.
\newblock {\em Neuron}, 77(1):141--154.

\bibitem[Peyrache and Buzs{\'a}ki, 2015]{peyrache2015th1}
Peyrache, A. and Buzs{\'a}ki, G. (2015).
\newblock Extracellular recordings from multi-site silicon probes in the
  anterior thalamus and subicular formation of freely moving mice
  (http://dx.doi.org/10.6080/k0g15xs1).

\bibitem[Peyrache et~al., 2015]{peyrache2015internally}
Peyrache, A., Lacroix, M.~M., Petersen, P.~C., and Buzs{\'a}ki, G. (2015).
\newblock Internally organized mechanisms of the head direction sense.
\newblock {\em Nature neuroscience}, 18(4):569--575.

\bibitem[Peyrache et~al., 2017]{peyrache2017transformation}
Peyrache, A., Schieferstein, N., and Buzs{\'a}ki, G. (2017).
\newblock Transformation of the head-direction signal into a spatial code.
\newblock {\em Nature communications}, 8(1):1752.

\bibitem[Rieke et~al., 1993]{rieke1993coding}
Rieke, F., Warland, D., and Bialek, W. (1993).
\newblock Coding efficiency and information rates in sensory neurons.
\newblock {\em EPL (Europhysics Letters)}, 22(2):151.

\bibitem[Roudi and Moser, 2014]{roudi2014grid}
Roudi, Y. and Moser, E.~I. (2014).
\newblock Grid cells in an inhibitory network.
\newblock {\em nature neuroscience}, 17(5):639--641.

\bibitem[Russo and Durstewitz, 2017]{russo2017cell}
Russo, E. and Durstewitz, D. (2017).
\newblock Cell assemblies at multiple time scales with arbitrary lag
  constellations.
\newblock {\em Elife}, 6:e19428.

\bibitem[Sargolini et~al., 2006]{sargolini2006conjunctive}
Sargolini, F., Fyhn, M., Hafting, T., McNaughton, B.~L., Witter, M.~P., Moser,
  M.-B., and Moser, E.~I. (2006).
\newblock Conjunctive representation of position, direction, and velocity in
  entorhinal cortex.
\newblock {\em Science}, 312(5774):758--762.

\bibitem[Sharp and Green, 1994]{sharp1994spatial}
Sharp, P.~E. and Green, C. (1994).
\newblock Spatial correlates of firing patterns of single cells in the
  subiculum of the freely moving rat.
\newblock {\em Journal of Neuroscience}, 14(4):2339--2356.

\bibitem[Shinomoto et~al., 2009]{shinomoto2009relating}
Shinomoto, S., Kim, H., Shimokawa, T., Matsuno, N., Funahashi, S., Shima, K.,
  Fujita, I., Tamura, H., Doi, T., Kawano, K., et~al. (2009).
\newblock Relating neuronal firing patterns to functional differentiation of
  cerebral cortex.
\newblock {\em PLoS computational biology}, 5(7):e1000433.

\bibitem[Shinomoto et~al., 2005]{shinomoto2005measure}
Shinomoto, S., Miura, K., and Koyama, S. (2005).
\newblock A measure of local variation of inter-spike intervals.
\newblock {\em Biosystems}, 79(1-3):67--72.

\bibitem[Shinomoto et~al., 2003]{shinomoto2003differences}
Shinomoto, S., Shima, K., and Tanji, J. (2003).
\newblock Differences in spiking patterns among cortical neurons.
\newblock {\em Neural Computation}, 15(12):2823--2842.

\bibitem[Skaggs et~al., 1993]{skaggs1993information}
Skaggs, W.~E., McNaughton, B.~L., and Gothard, K.~M. (1993).
\newblock An information-theoretic approach to deciphering the hippocampal
  code.
\newblock In {\em Advances in neural information processing systems}, pages
  1030--1037.

\bibitem[Skaggs et~al., 1996]{skaggs1996theta}
Skaggs, W.~E., McNaughton, B.~L., Wilson, M.~A., and Barnes, C.~A. (1996).
\newblock Theta phase precession in hippocampal neuronal populations and the
  compression of temporal sequences.
\newblock {\em Hippocampus}, 6(2):149--172.

\bibitem[Solstad et~al., 2008]{solstad2008representation}
Solstad, T., Boccara, C.~N., Kropff, E., Moser, M.-B., and Moser, E.~I. (2008).
\newblock Representation of geometric borders in the entorhinal cortex.
\newblock {\em Science}, 322(5909):1865--1868.

\bibitem[Song et~al., 2018]{song2017emergence}
Song, J., Marsili, M., and Jo, J. (2018).
\newblock Resolution and relevance trade-offs in deep learning.
\newblock {\em Journal of Statistical Mechanics: Theory and Experiment},
  2018(12):123406.

\bibitem[Stein, 1967]{stein1967information}
Stein, R.~B. (1967).
\newblock The information capacity of nerve cells using a frequency code.
\newblock {\em Biophysical journal}, 7(6):797--826.

\bibitem[Stein et~al., 2005]{stein2005neuronal}
Stein, R.~B., Gossen, E.~R., and Jones, K.~E. (2005).
\newblock Neuronal variability: noise or part of the signal?
\newblock {\em Nature Reviews Neuroscience}, 6(5):389.

\bibitem[Stensola et~al., 2012]{stensola2012entorhinal}
Stensola, H., Stensola, T., Solstad, T., Fr{\o}land, K., Moser, M.-B., and
  Moser, E.~I. (2012).
\newblock The entorhinal grid map is discretized.
\newblock {\em Nature}, 492(7427):72--78.

\bibitem[Stensola et~al., 2015]{stensola2015shearing}
Stensola, T., Stensola, H., Moser, M.-B., and Moser, E.~I. (2015).
\newblock Shearing-induced asymmetry in entorhinal grid cells.
\newblock {\em Nature}, 518(7538):207--212.

\bibitem[Strong et~al., 1998]{strong1998entropy}
Strong, S.~P., Koberle, R., van Steveninck, R. R. d.~R., and Bialek, W. (1998).
\newblock Entropy and information in neural spike trains.
\newblock {\em Physical review letters}, 80(1):197.

\bibitem[Taube, 1995]{taube1995head}
Taube, J.~S. (1995).
\newblock Head direction cells recorded in the anterior thalamic nuclei of
  freely moving rats.
\newblock {\em Journal of Neuroscience}, 15(1):70--86.

\bibitem[Taube et~al., 1990]{taube1990head}
Taube, J.~S., Muller, R.~U., and Ranck, J.~B. (1990).
\newblock Head-direction cells recorded from the postsubiculum in freely moving
  rats. i. description and quantitative analysis.
\newblock {\em Journal of Neuroscience}, 10(2):420--435.

\bibitem[Treves and Panzeri, 1995]{treves1995upward}
Treves, A. and Panzeri, S. (1995).
\newblock The upward bias in measures of information derived from limited data
  samples.
\newblock {\em Neural Computation}, 7(2):399--407.

\bibitem[Zhang et~al., 1998]{zhang1998interpreting}
Zhang, K., Ginzburg, I., McNaughton, B.~L., and Sejnowski, T.~J. (1998).
\newblock Interpreting neuronal population activity by reconstruction: unified
  framework with application to hippocampal place cells.
\newblock {\em Journal of neurophysiology}, 79(2):1017--1044.

\end{thebibliography}

% this is to relabel the equation numbers
\newcounter{defcounter}
\setcounter{defcounter}{0}
\newenvironment{myequation}{%
\addtocounter{equation}{-1}
\refstepcounter{defcounter}
\renewcommand\theequation{S\thedefcounter}
\begin{equation}}
{\end{equation}}

% this is to relabel the figures
\setcounter{figure}{0}
\renewcommand{\thefigure}{S\arabic{figure}}

\renewcommand{\thesection}{Text S\arabic{section}}

\newcommand{\fix}{\marginpar{FIX}}
\newcommand{\new}{\marginpar{NEW}} 
\renewcommand\Authfont{\fontsize{12}{10}\selectfont}
\newcommand{\etalp}{\textit{et. al.}}
\newcommand{\etal}{\textit{et. al. }}

\afterpage{%
\appendix
\centering
\vspace*{\fill}
\section{Supplementary Materials}
\vspace*{\fill}

\begin{figure}[!h]
\begin{center}
\includegraphics[width=0.77\textwidth]{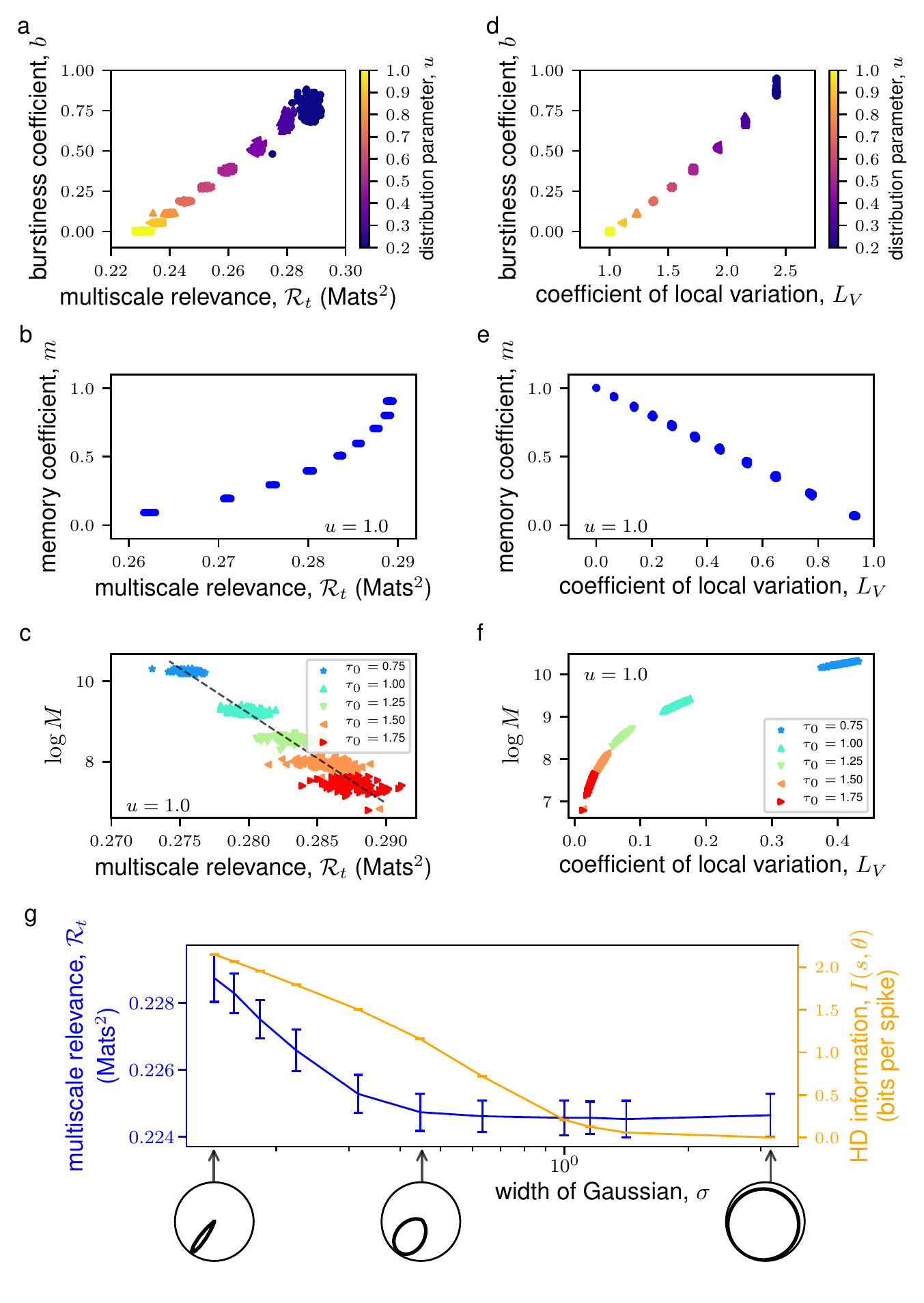}
%\caption{{\bf Synthetically-generated neural data reveals relationship of the MSR and of the coefficient of local variation, $L_V$, with the bursty-ness and memory coefficients.}
%Interevent times were drawn from a stretched exponential distribution to simulate random events up to 100,000 time units where short-term memory effects were introduced through a shuffling procedure and the number of random events, $M$, were varied by modifying the characteristic time constant, $\tau_0$ (See Supplementary Text S1 for details).
%Scatter plots show how the multiscale relevance (MSR) scales with the bursty-ness coefficient, $b$ (panel {\bf a}), the memory coefficient, $m$ (panel {\bf b}), and $\log M$ (panel {\bf c}).
%In panel {\bf b}, random events were drawn from a stretched exponential distribution with $u=1.0$ while in panel {\bf c}, the parameter $u$ was set to $0.3$.
%Panels {\bf d}, {\bf e} and {\bf f}, on the other hand, show the relationship between $L_V$ and bursty-ness coefficient, memory coefficient and $\log M$ respectively.
%The results for 100 realizations of such random events are shown.
%Notice, in {\bf c} and {\bf f}, that both the MSR and the $L_V$ are sensitive to the number of spiking events.}

\caption{\label{figure FigureS1}
{\bf Synthetically-generated neural data reveals relationship of the MSR and of the coefficient of local variation, $L_V$, with the bursty-ness and memory coefficients, and Skaggs-McNaughton mutual information.}
Interevent times were drawn from a stretched exponential distribution to simulate random events up to 100,000 time units where short-term memory effects were introduced through a shuffling procedure and the number of random events, $M$, were varied by modifying the characteristic time constant, $\tau_0$ (See Supplementary Text S1 for details).
Scatter plots show how the multiscale relevance (MSR) scales with the bursty-ness coefficient, $b$ (panel {\bf a}), the memory coefficient, $m$ (panel {\bf b}), and $\log M$ (panel {\bf c}).
In panel {\bf b}, random events were drawn from a stretched exponential distribution with $u=1.0$ while in panel {\bf c}, the parameter $u$ was set to $0.3$.
Panels {\bf d}, {\bf e} and {\bf f}, on the other hand, show the relationship between $L_V$ and bursty-ness coefficient, memory coefficient and $\log M$ respectively.
The results for 100 realizations of such random events are shown.
Notice, in {\bf c} and {\bf f}, that both the MSR and the $L_V$ are sensitive to the number of spiking events.
Idealized HD cells were also simulated by assuming that the firing probability conditioned on the HD follows a circular Gaussian distribution centered at a preferred HD, $\theta_0$, with a given width, $\sigma$. Spike train data was generated using resampling methods (See Section 5.5) using a random walk trajectory for a fixed $\theta_0$ and $\sigma$. For each realization of the spike train data, the MSR (blue line) and the HD information (orange line) were calculated and shown in panel {\bf g}. Here, we report the mean and standard deviation across 100 spike train realization. For a given $\theta_0$, sample HD tuning curves are shown in the polar plots.}
\end{center}
\end{figure}

\begin{figure}
\begin{center}
\includegraphics[width=\textwidth]{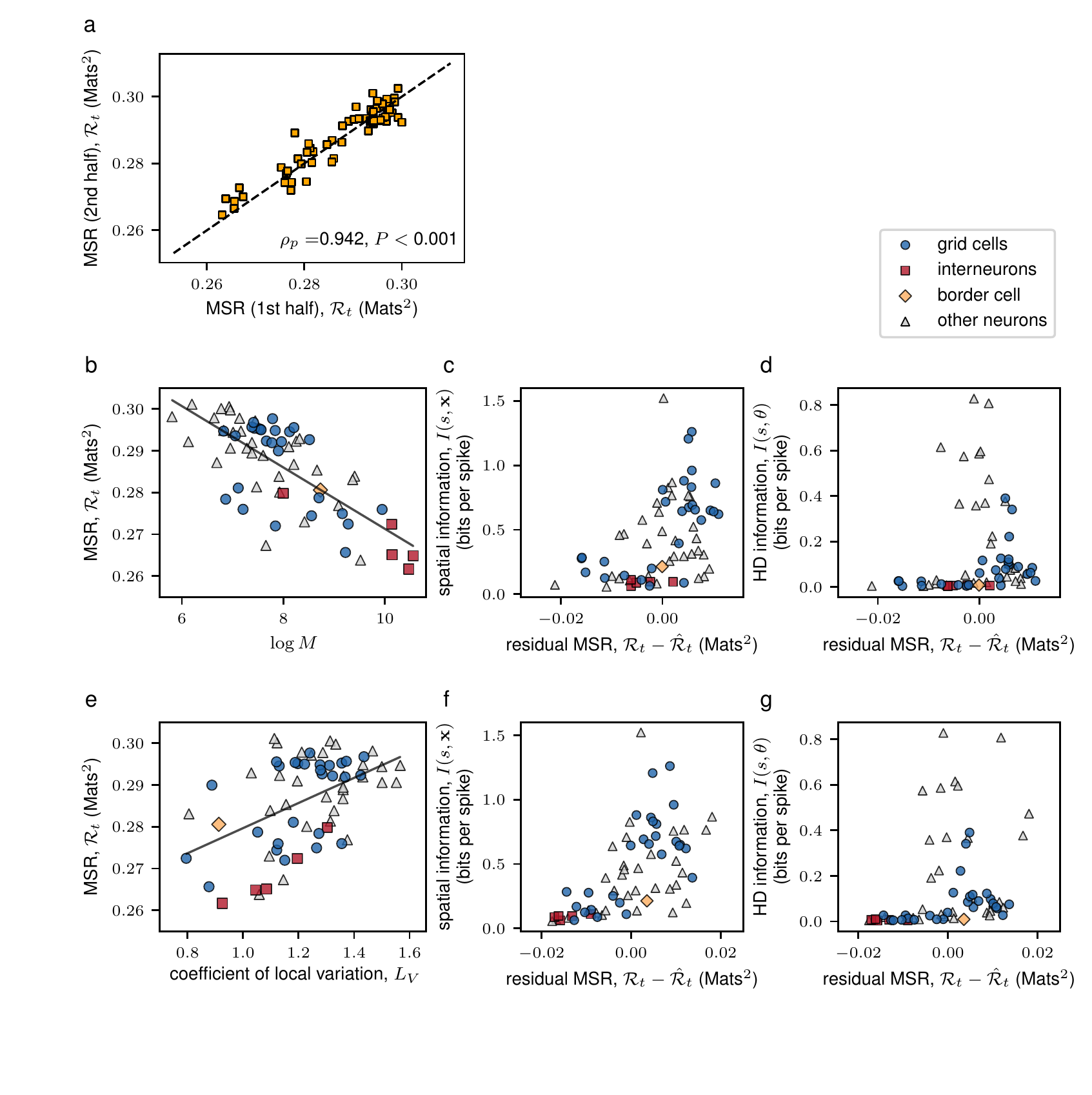}
\caption{\label{figure FigureS2}
{\bf The MSR is a robust measure and contains information beyond what the number of spikes and local variations can explain.}
For each neuron, the MSR was calculated using only the first half and only the second half of the data  (panel {\bf a}).
The scatter plot reports the two results. 
The linearity of the relationship between the two sets of partial data is quantified by the Pearson correlation $\rho_p$ along with its $P$-value.
The black dashed line indicates the linear fit.
For the neurons in the mEC dataset, the MSR was linearly regressed with $\log M$ (panel {\bf b}).
The residual MSR, defined as the deviation of the MSR from the black regression line, were then correlated against spatial (panel {\bf c}) and HD (panel {\bf d}) information.
The MSR was also linearly regressed with the coefficient of local variation, $L_V$ (panel {\bf e}).
The residual MSRs were then correlated against spatial (panel {\bf f}) and HD (panel {\bf g}) information.}
\end{center}
\end{figure}

\begin{figure}
\begin{center}
\includegraphics[width=\textwidth]{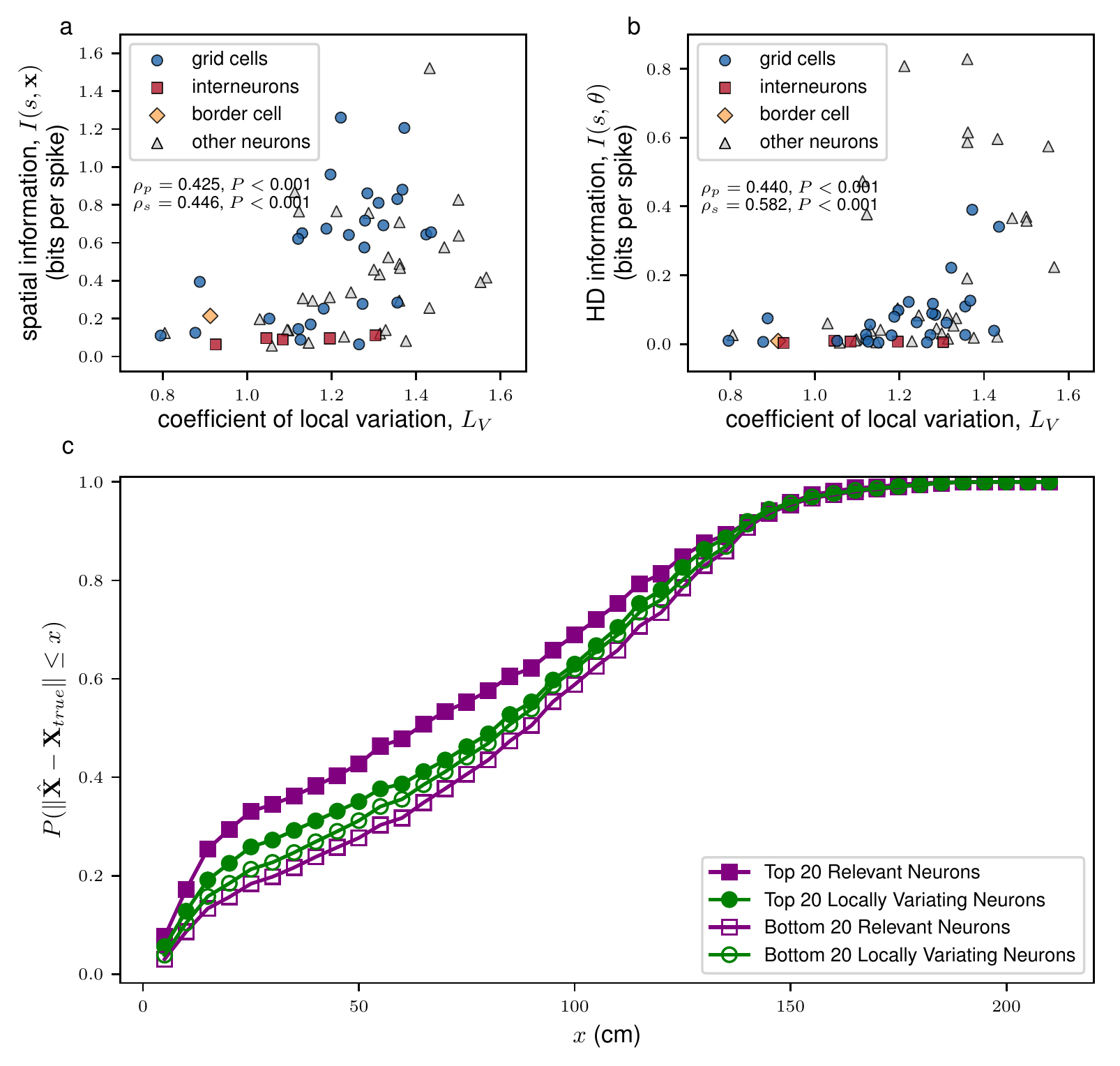}
\caption{\label{figure FigureS3}
{\bf Local variations in the interspike intervals can capture spatial and HD information but not decodable spatial information.}
A scatter plot of the coefficient of local variation $L_V$ vs. the spatial (HD) information is shown in {\bf a} ({\bf b}).
The shapes of the scatter points indicate the identity of the neuron according to \citet*{stensola2012entorhinal}.
The linearity and monotonicity of the multiscale relevance and the information measures were assessed by the Pearson's correlation, $\rho_p$, and the Spearman's correlation, $\rho_s$, respectively.
The 20 top and bottom locally variating neurons (LVNs) were then used to decode position (See Main Text Section 5.6).
Panel {\bf c} shows the cumulative distribution of the decoding error, $\| \hat{\mathrm{\mathbf X}} - {\mathrm{\mathbf X}}_{true} \|$, for the RNs (solid violet squares) and LVNs (solid green circles) neurons as well as for the non-RNs (dashed violet squares) and non-LVNs (dashed green circles).
In all the decoding procedures, time points where all the neurons in the ensemble was silent were discarded in the decoding process.}
\end{center}
\end{figure}

\begin{figure}
\begin{center}
\includegraphics[width=\textwidth]{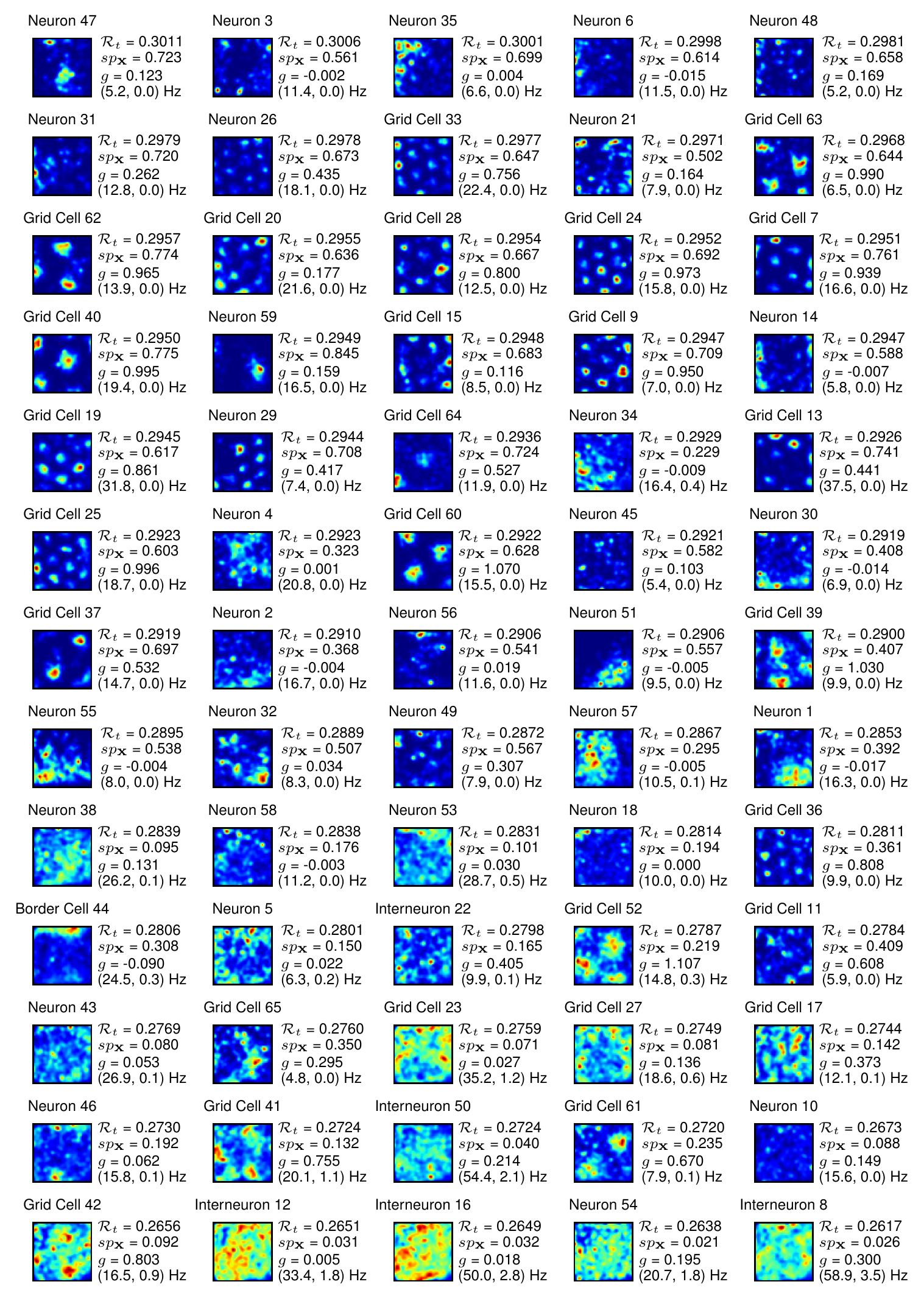}
\caption{\label{figure FigureS4}
{\bf RNs in the mEC exhibit spatially selective firing compared to non-RNs.}
The spatial firing rate maps of the 65 neurons in the mEC data, sorted according to their MSR scores, are shown together with the calculated spatial sparsity, $sp_{\mathrm{\mathbf x}}$, the grid score, $g$, and the maximum and minimum firing.}
\end{center}
\end{figure}

\begin{figure}
\begin{center}
\includegraphics[width=\textwidth]{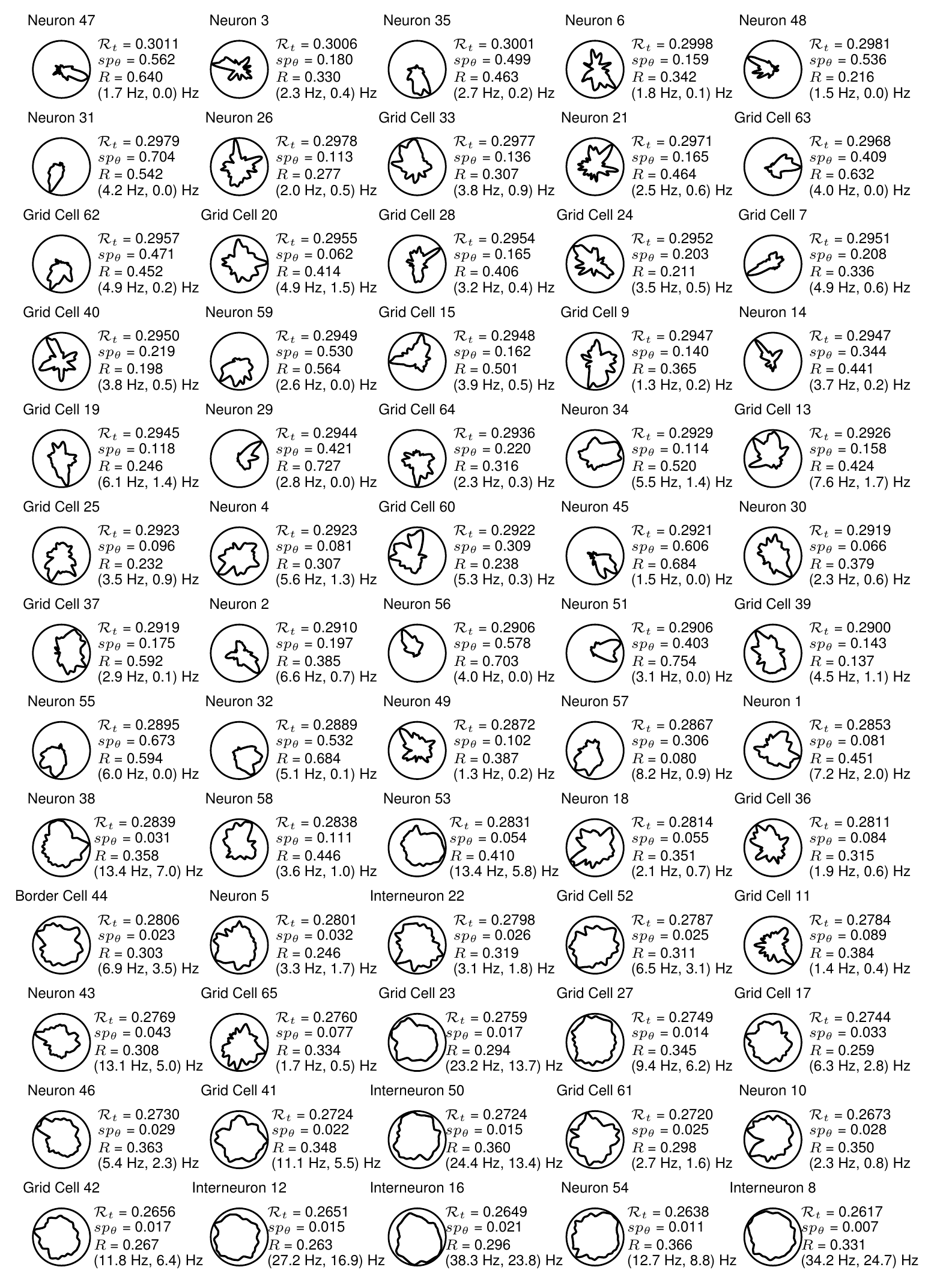}
\caption{\label{figure FigureS5}
{\bf RNs in the mEC exhibit HD selective firing compared to non-RNs.}
The HD tuning curves of the 65 neurons in the mEC data, sorted according to their MSR scores, are shown together with the calculated HD sparsity, $sp_\theta$, the Rayleigh mean vector length, $R$, and the maximum and minimum firing.}
\end{center}
\end{figure}
\clearpage

\begin{figure}
\begin{center}
\includegraphics[width=\textwidth]{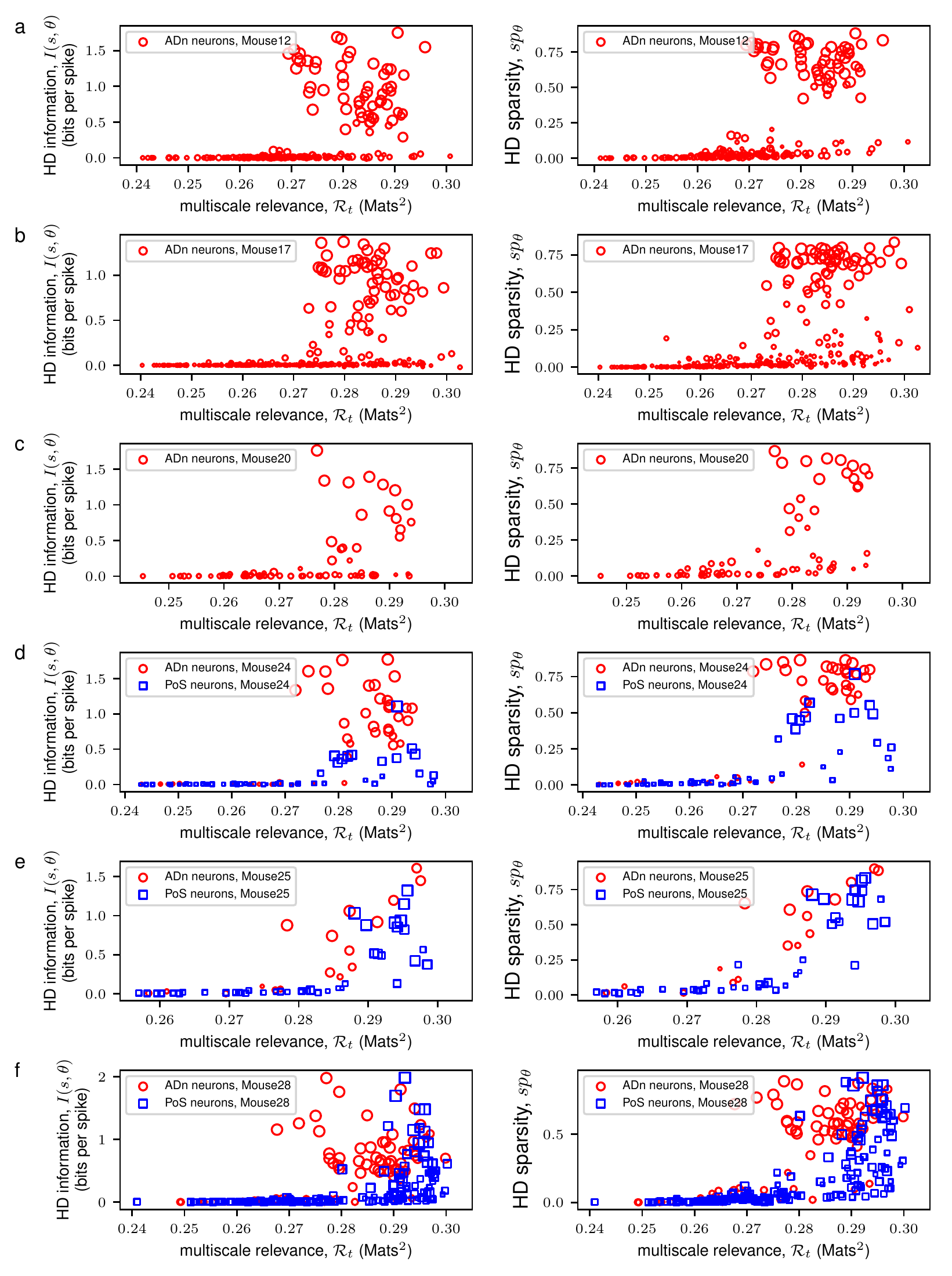}
\caption{\label{figure FigureS6}
{\bf MSR of neurons from the anterodorsal thalamic nucleus (ADn) and post-subicular (PoS) regions of 6 freely-behaving mice pooled from multiple recording sessions}.
For each mouse, the MSR of the recorded neurons which had more than 100 recorded spikes in a session were calculated.
The corresponding the HD information and sparsity (in bits per spike, see Main Text Section 5.4: Information, Sparsity and other Scores) were also calculated.
ADn neurons are depicted in red circles while PoS neurons in blue squares.
The size of each point reflects the mean vector lengths of the neurons wherein larger points indicate a unimodal distribution in the calculated HD tuning curves.}
\end{center}
\end{figure}

\begin{figure}
\begin{center}
\includegraphics[width=\textwidth]{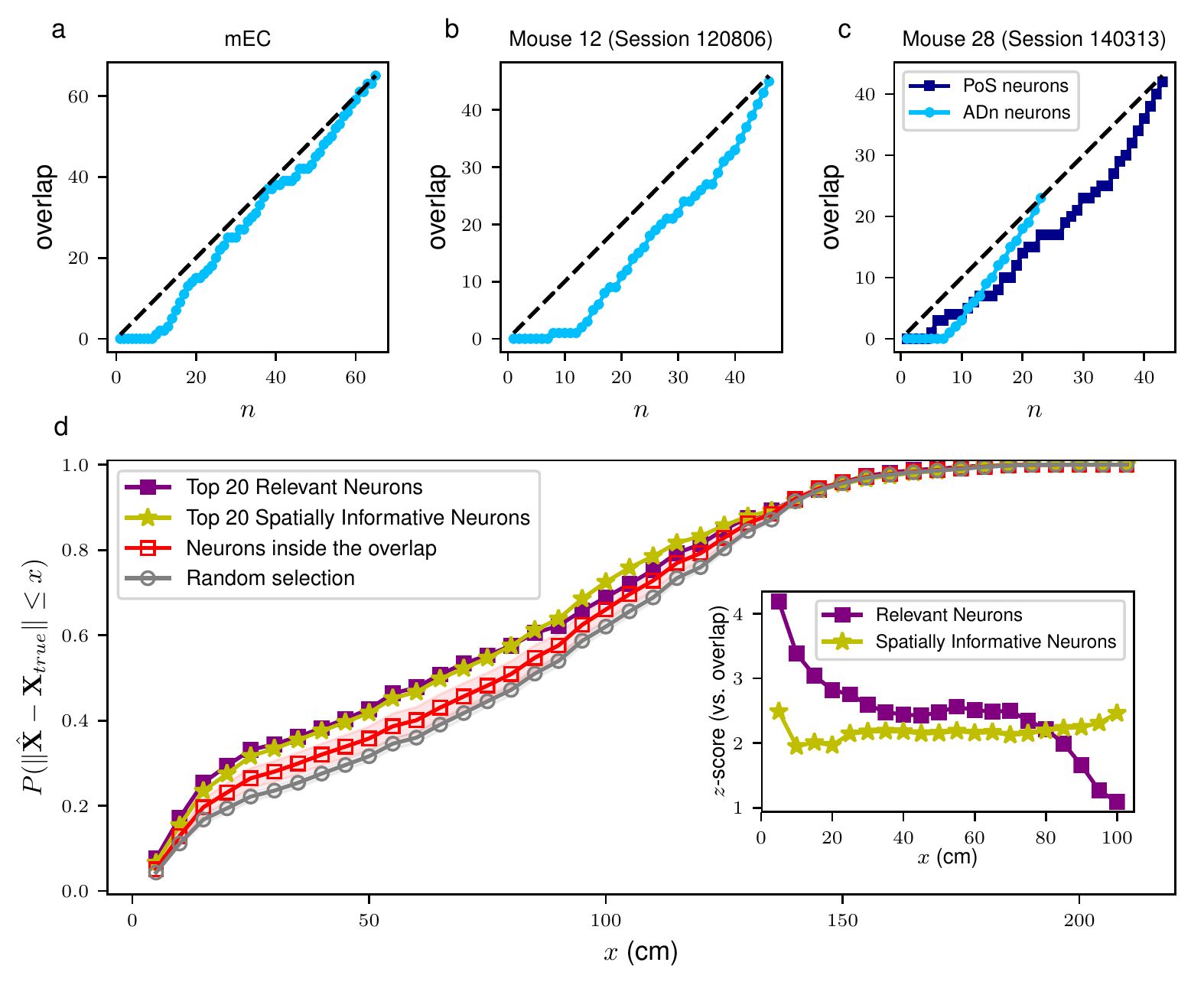}
\caption{\label{figure FigureS7}
{\bf The mEC neurons that were both RNs and INs do not contain the bulk of the decodable spatial information.}
The overlap between the set of RNs and INs as a function of the size, $n$, of each set for the mEC ({\bf a}), for the ADn of Mouse 12 (Session 120806) ({\bf b}) and for the ADn and PoS of Mouse 28 (Session 140313) ({\bf c}).
To show how much decodable spatial information there is in the overlap between the RNs and spatial INs in the mEC at $n=20$, we took the 14 overlapping neurons (ONs) and randomly chose 6 neurons outside of this overlap and performed a Bayesian positional decoding (see Main Text Section 5.6).
The mean and standard errors of the cumulative distribution of decoding errors, $\| \hat{\mathrm{\mathbf X}} - {\mathrm{\mathbf X}}_{true} \|$, of the 14 ONs + 6 random neurons (n = 100 realizations) are shown in grey ({\bf D}) together with the cumulative distribution of decoding errors of the RNs (violet squares) and spatial INs (yellow stars).
For a given position error, $x$, a $z$-score can be calculated by measuring how many standard errors from the mean of the decoding errors for ONs is the decoding error of the RNs or of the spatial INs.
These $z$-scores are shown in the inset of panel {\bf d}.}
\end{center}
\end{figure}
\clearpage

\begin{figure}
\begin{center}
\includegraphics[width=\textwidth]{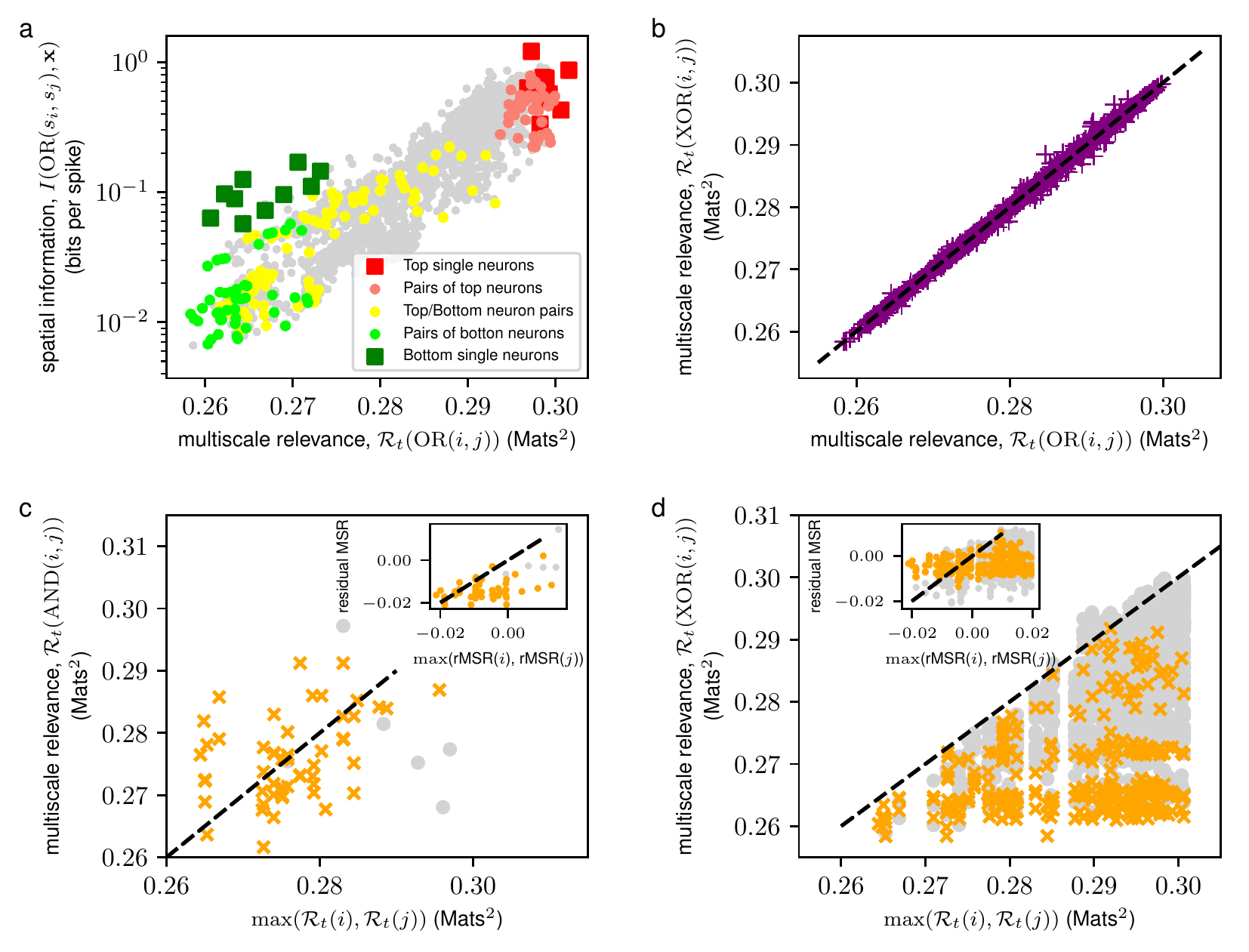}
\caption{\label{figure FigureS8}
{\bf Multiscale relevance (MSR) for spike trains built from AND, OR and XOR Boolean functions between pairs of neurons.} To study the possible application of MSR to neuronal ensembles,
MSRs were calculated from spike trains resulting from AND, OR and XOR Boolean function between pairs of neurons (See Section 5.7). A scatter plot between the spatial information and the calculated MSR for spike trains constructed from OR Boolean function is shown in {\bf a}. Red squares indicate the 10 single neurons with the highest MSR (top single neurons), green squares the 10 single neurons with the lowest MSRs (bottom single neurons), red circles the OR Boolean function from pairs of top neurons, green circles the pairs of bottom neurons, and yellow circles the pairs of top and bottom neurons. Similar results to panel {\bf a} hold for HD information (plot not shown). Panel {\bf b} shows a scatter plot between MSRs from OR Boolean function and from XOR Boolean function.
The inset scatter plot show the difference between the MSRs from XOR and from OR Boolean function as a function of the average firing rate of the neuron pairs along with the best-fit line (dashed line).
The comparison between the MSR of the OR and the XOR potentially allows us to probe the effects of common driving of the neurons, because the latter includes also simultaneous firing, whereas the former does not. We find that the MSR of the XOR is very similar to that of the OR. One way to interpret this lack of difference is that the role of common input to correlated firing is small. The lack of difference could also be due to small firing rates of neurons but given that we see the similarity between the MSR of the OR and XOR for almost all pairs makes this unlikely.
Panel {\bf c} ({\bf d}) shows a scatter plot between the  MSRs from an AND (XOR) Boolean function between pairs of neurons and the maximal MSR between the corresponding pairs. Residual MSRs are calculated by subtracting the part of the MSR that is explained by $\log M$ through linear regression. The corresponding inset scatter plots show a scatter plot between residual MSRs from pairs of neurons and the maximum residual MSR of the corresponding pairs. Orange crosses indicate neurons that are paired with an interneuron.
The resulting MSR (of the OR or of the XOR) is almost always lower (and never significantly higher) than the MSR of the most relevant of the two neurons, suggests that individual neurons contain non-redundant information.
Furthermore, for those pairs of neurons where we get a significant signal, in most of the cases, one of the neurons turns out to be an interneuron which suggests that interneurons play a peculiar role in the information aggregation.
The neuron pairs considered in Fig. \ref{figure FigureS9} are highlighted. GC: grid cell, I: interneuron, N: unclassified neuron.}
\end{center}
\end{figure}
\clearpage

\begin{figure}
\begin{center}
\includegraphics[width=\textwidth]{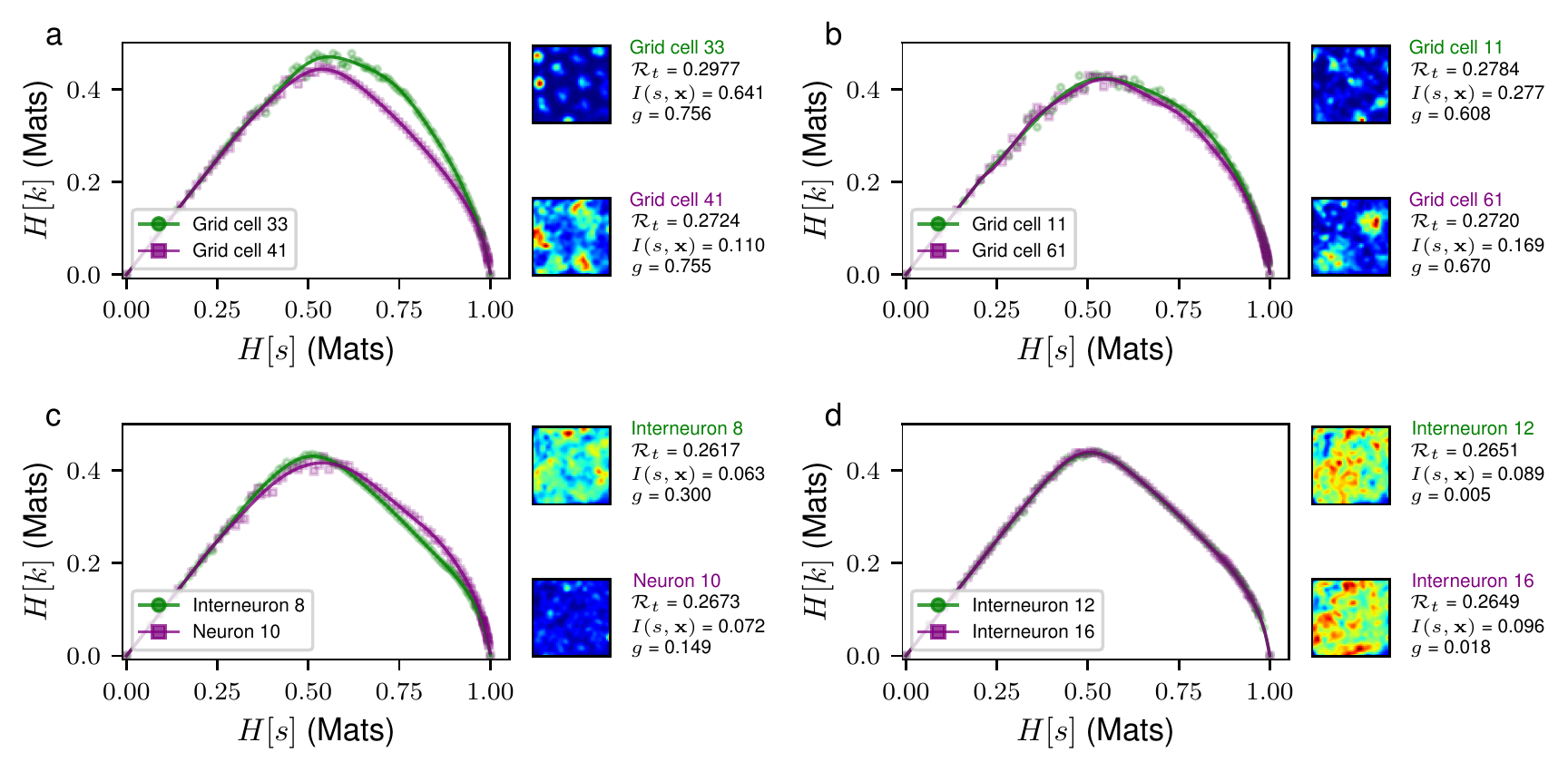}
\caption{\label{figure FigureS9}
{\bf Resolution vs. relevance curves for different pairs of neurons.} Panels show the resolution, $H[s]$, vs relevance $H[k]$ curves traced by different pairs of neurons: grid cells with high and low MSR ({\bf a}), grid cells with low MSRs having similar grid scores ({\bf b}), a neuron having a low MSR and an interneuron ({\bf c}) and two interneurons ({\bf d}). Each point, $(H[s], H[K])$, in this curve corresponds to a fixed binning time, $\Delta t$, with which we see the corresponding spike count codes. 
The spatial noise correlation, $C_{ij}(\mathrm{\mathbf x})$ (see main text, section 5.8), for the neuron pairs are also indicated. The multiscale relevance of the boolean functions that can be constructed from these neuron pairs are indicated in Fig. \ref{figure FigureS8}.}
\end{center}
\end{figure}
\clearpage
}

\afterpage{
\section*{Text S1: Relation between MSR and other measures of temporal structure}\label{Text S1}
Characterizing the neural spiking can be done by studying the distribution of the time intervals between two succeeding spikes, known in literature as the interspike interval (ISI) distribution which allows us to see whether a neuron fires in bursts~\citep*{ebbesen2016cell, sharp1994spatial}.
Note that given the time stamps of neural activity $\lbrace t_1, \ldots, t_M \rbrace$, the interspike interval is given by $\lbrace \tau_1, \ldots, \tau_{M-1} \rbrace$ where $\tau_i = t_{i+1} - t_i$.
Because the multiscale relevance (MSR) is built to separate relevant neurons from the irrelevant ones through their temporal structures in the neural spiking, we wanted to assess how the proposed measure scales with the characteristics that give structure to temporal events.
In the context of the temporal activity of a neuron, a feature of the relevance measure, $H[K]$ is that highly regular, equally-spaced ISI are attributed with a low measure.
On the other hand, ISI that follow broad, non-trivial distributions are attributed with a high relevance measure.
Hence, we expected that the relevance measure, and therefore the MSR, captures non-trivial bursty patterns of neurons.

To study how MSR behaves with respect to the characteristics of ISI, we considered a stretched exponential distribution
\begin{myequation}
P_{SE}^u(\tau) = \frac{u}{\tau_0} \left[ \frac{\tau}{\tau_0} \right]^{u-1} \exp \left[ - \left( \frac{\tau}{\tau_0} \right)^u \right]
\label{stretched exponential}
\end{myequation}
with which the parameter $u$ allows us to define the broadness of underlying distribution and $\tau_0$ is the characteristic time constant of the random event.
For Poisson processes, the ISI follow an exponential distribution corresponding to $u=1$ in Eq. \eqref{stretched exponential}.
For $u<1$, the ISI distribution becomes broad and tends to a power law distribution with an exponent of $-1$ in the limit when $u \to 0$.
On the other hand, for $u>1$, the distribution becomes narrower and tends to a Dirac delta function in the limit when $u \to \infty$.

Upon fixing the parameters $u$ and $\tau_0$ which fixes the stretched exponential distribution in Eq. \eqref{stretched exponential}, random ISI, $\tau_i$, could then be sampled independently from Eq.~(\ref{stretched exponential}) so as to generate a time series of 
%using the inverse transform method where we simulate up to 
100,000 time units.
%To do this, we first calculated the cumulative distribution, $F_{SE}^u(\tau)$, of Eq. \eqref{stretched exponential} given by
%\begin{equation}
%F_{SE}^u(\tau) = 1 - \exp \left[ - \left( \frac{\tau}{\tau_0} \right)^u \right].
%\end{equation}
%The inverse transform method, then, takes advantage of the theorem that a random number, $q \in [0,1]$, drawn from the cumulative distribution, $F_{SE}^u(\tau)$, is uniformly distributed~\cite{feller1968introduction}. Once a uniform random number, $q$, is generated, one can then invert the relationship $q=F_{SE}^u(\tau)$ to solve for $\tau$~\cite{devroye1986introduction}. In our case, we stopped drawing uniform random numbers whenever $\sum_i \tau_i \geq 100,000$ time units.
The MSRs of each time series could then be calculated using the methods described in the main text (Section 2).

To characterize the temporal structures of both the simulated data and neural data, we adapted the measures of bursty-ness and memory of Goh and Barabasi \citep*{goh2008burstiness}.
While the bursty-ness coefficient, $b$ defined as
\begin{myequation}
b = \frac{\sigma_{\tau} - \mu_{\tau}}{\sigma_{\tau} + \mu_{\tau}},
\label{burstiness coefficient}
\end{myequation}
measures the broadness of the underlying ISI distribution with $\mu_\tau$ and $\sigma_\tau$ as the mean and standard deviations of the ISI respectively, the memory coefficient, $m$ defined as
\begin{myequation}
m = \frac{1}{M - 2} \sum_{j=1}^{M - 2} \frac{(\tau_j - \mu_{\tau})(\tau_{j+1} - \mu_{\tau})}{\sigma_{\tau}^2},
\label{memory coefficient}
\end{myequation}
measures the short-time correlation between events.

For the stretched exponential distribution in Eq. \eqref{stretched exponential}, the mean and standard deviations could be computed as
\begin{myequation}
\mu_\tau = \tau_0 \Gamma \left( \frac{u+1}{u} \right)
\end{myequation}
and
\begin{myequation}
\sigma_\tau = \tau_0 \sqrt{ \Gamma \left( \frac{u+2}{u} \right) - \Gamma \left( \frac{u+1}{u} \right)^2 }
\end{myequation}
where $\Gamma(x) \equiv (x-1)!$ is the gamma function.
With these closed-form relationships, we could now study the limiting properties of the burstiness and memory coefficients.
For Poisson processes, the mean, $\mu_\tau$, and standard deviation, $\sigma_\tau$, coincide, i.e. $\mu_\tau = \sigma_\tau = \tau_0$,  and thus with Eq. \eqref{burstiness coefficient}, give $b=0$.
For broad distributions, $u<1$ in Eq. \eqref{stretched exponential}, $\sigma_\tau > \mu_\tau$ which gives $b>0$ and tends to approach $b \to 1$ in the limit $u \to 0$. On the other hand, for narrow distributions, $u>1$ in Eq. \eqref{stretched exponential}, $\sigma_\tau < \mu_\tau$ resulting to $b<0$ and tends to $b \to -1$ in the limit $u \to \infty$.
Hence, the bursty-ness parameter, $b$, is a bounded parameter, i.e., $b \in [-1, 1]$.

For the synthetic datasets, note that fixing the parameter $u$ automatically fixes the bursty-ness coefficient, $b$.
However, because the synthetic ISI are sampled independently, the memory coefficient, $m$, is approximately zero.
Short-term memory can then be introduced by first sorting the ISI in decreasing (or increasing) order which results to $m \approx 1$.
Randomly shuffling a subset of the ordered ISI (100 events at a time in this case) results to a monotonic decrease of $m$.
In the limit of infinite data, the memory coefficient is bounded by $[-1, 1]$.
These bounds may no longer hold in the case of limited data. Despite this, a positive memory coefficient indicates that a short (long) ISI between events tends to be followed by another short (long) interval and a negative memory coefficient indicates that a short (long) ISI between events tends to be followed by a long (short) interval. 

With this, we found that the MSR increased with bursty-ness and memory for the synthetically generated dataset as seen in Fig. \ref{figure FigureS1}a and b.
We also sought to characterize the relationship between the number of events, $M$, with the MSR which can be addressed by changing the characteristic time constant, $\tau_0$, in Eq. \eqref{stretched exponential} wherein decreasing $\tau_0$ leads to more events and thus, increased $\log M$.
We found that MSR decreased with $\log M$ as seen in Fig. \ref{figure FigureS1}c.
This result is indicative that MSR of randomly generated events can be explained by $\log M$.

Since the MSR is constructed as a measure of dynamical variablity, we compared our results on synthetically generated datasets with the coefficient of local variation, $L_V$,~\citep*{shinomoto2005measure,shinomoto2003differences,shinomoto2009relating} defined as
\begin{equation}
L_V = \frac{1}{M-1}\sum_{j=1}^{M-1} \frac{3 \left( \tau_j - \tau_{j+1} \right)^2}{\left( \tau_j + \tau_{j+1} \right)^2}
\end{equation}
where the factor 3 in the summand was taken such that, for a Poisson process, $L_V = 1$.
With this, we found that the $L_V$ increases with increasing bursty-ness coefficient, $b$, indicating that power law ISI distributions lead to highly locally variating spiking events (see Fig. \ref{figure FigureS1}d).
Also, we found that the $L_V$ decreases with increasing short-term memory, $m$ (see Fig. \ref{figure FigureS1}e).
Finally, like the MSR, we also found a dependence of the $L_V$ with the $\log M$ (see Fig. \ref{figure FigureS1}f).

Following the results on synthetic data, we also analyzed temporal characteristics in real neural dataset.
In the case of neurons in the mEC data, we also found that MSR decreased with the logarithm of the number of observed spikes, $\log M$, as shown in Fig. \ref{figure FigureS2}b.
To determine how much of the calculated MSRs can be explained by the number of observed spikes, $M$, we linearly regressed MSR with $\log M$ shown as the dashed line in Fig. \ref{figure FigureS2}b.
Residuals were then calculated as the deviation of the calculated MSR from the regression line and thus, captures the amount of MSR that cannot be explained by $\log M$ alone.
We showed in Fig. \ref{figure FigureS2}c and d that the MSR for real dataset still contained information going beyond $\log M$ as the residual MSRs (with respect to $\log M$) still retained the dependence with spatial and HD information as already observed in the main text (Fig. 2).
We also observed a positive correlation between MSR and $L_V$.
However, through residual analysis, we found that the residual MSRs (with respect to $L_V$) still contained spatial and HD information as seen in Fig. \ref{figure FigureS2}f and g.}

\end{document}